\documentclass{nature}
\input{epsf}

\usepackage{amssymb}
\usepackage[dvips]{graphicx}
\usepackage{graphics}
\usepackage{txfonts}
\usepackage{times,epsfig}
\usepackage{caption}

\usepackage{natbib} 
\usepackage{multibib}
\newcites{meth}{ }
\newcites{sup}{ }

\bibpunct[, ]{(}{)}{,}{n}{}{,}

\newcommand{\Msun}{\mathrm{M}_{\odot}}

\newcommand{\Lsun}{\mathrm{L}_{\odot}}
\newcommand{\ud}{\mathrm{d}}

\title{Rapid formation of large dust grains in the luminous \\
supernova SN 2010jl} 
\author{Christa Gall$^{1,2,3}$, Jens Hjorth$^{2}$, Darach Watson$^{2}$, Eli Dwek$^{3}$, Justyn R. Maund$^{4,2}$, Ori Fox$^{5}$,  Giorgos Leloudas$^{6,2}$, Daniele Malesani$^{2}$ \& Avril C. Day-Jones$^{7}$}
\begin{document}

\maketitle

%
\begin{affiliations}
\item Department of Physics and Astronomy, Aarhus University, Ny Munkegade 120, DK-8000 Aarhus C, Denmark
\item Dark Cosmology Centre, Niels Bohr Institute, University of Copenhagen, Juliane Maries Vej 30, DK-2100 Copenhagen \O, Denmark
\item Observational Cosmology Lab, Code 665, NASA Goddard Space Flight Center, Greenbelt, MD 20771, USA
\item Astrophysics Research Centre School of Mathematics and Physics Queen's University Belfast Belfast BT7 1NN, UK
\item Department of Astronomy, University of California, Berkeley, CA 94720-3411, USA
\item The Oskar Klein Centre, Department of Physics, Stockholm University, Albanova University Centre, 10691, Stockholm, Sweden
\item Departamento de Astronomia, Universidad de Chile, Camino del Observatorio 1515, Santiago, Chile
\end{affiliations}
%
%
\begin{abstract}
The origin of dust in galaxies is still a mystery
\citep{2011A&ARv..19...43G, 2009MNRAS.396..918M,
2009ASPC..414..453D, 2011MNRAS.417.1510D}. 
The majority of the refractory 
elements are produced in supernova explosions but it is 
unclear how and where dust grains condense and grow, and 
how they avoid destruction in the harsh environments of 
star-forming galaxies. The recent detection of 0.1--0.5 solar 
masses of dust in nearby supernova remnants 
\citep{2011Sci...333.1258M,2014ApJ...782L...2I,2012ApJ...760...96G} 
suggests {\em in situ} dust formation, while other observations reveal 
very little dust in supernovae the first few years after 
explosion \citep{2011A&ARv..19...43G, 2004MNRAS.352..457P,
2012ApJ...744...26O,2012ApJ...756..173S}. 
Observations of the bright SN 2010jl have been interpreted 
as pre-existing dust \citep{2011AJ....142...45A}, dust 
formation \citep{2012AJ....143...17S, 2013ApJ...776....5M} or 
no dust at all  \citep{2012AJ....144..131Z}. Here we report 
the rapid (40--240 days) formation of dust in its dense 
circumstellar medium. The wavelength dependent extinction 
of this dust reveals the presence of very large ($> 1\ \mu$m) 
grains, which are resistant to destructive processes  
\citep{2010ApJ...715.1575S}. 
At later times (500--900 days), the near-IR thermal emission 
shows an accelerated growth in dust mass, marking the 
transition of the supernova from a circumstellar- to an 
ejecta-dominated source of dust. This provides the link 
between the early and late dust mass evolution in 
supernovae with dense circumstellar media.
\end{abstract}
%
%

We observed the bright ($V\sim14$) and luminous 
($M_V\sim-20$) Type IIn SN 2010jl 
\citep{2010CBET.2532....1N} with 
the VLT/X-shooter spectrograph 
covering the wide wavelength range 0.3--2.5 $\mu$m. Peak 
brightness occurred on 2010 Oct 18.6 UT, and observations 
were made at 9 early epochs and at one late epoch, 26--239 
and 868 days past peak, respectively  
(Methods, Extended Data Table 1, Extended Data Figures 1--5).
%
%
Figure~1 shows the intermediate-width components of the 
hydrogen emission lines of H$\gamma$ at 
$\lambda$4340.472 and P$\beta$ at 
$\lambda$12818.072 and of the oxygen ejecta emission 
lines [\ion{O}{i}] $\lambda\lambda$6300.304, 6363.776 (rest 
frame). The emission profiles change with time, exhibiting a 
substantial depression of the red wings and a corresponding 
blueshift of the centroids of the lines (Extended Data Figure 6)
due to preferential 
extinction of the emission from the receding material on the 
far side of the supernova 
\citep{1989LNP...350..164L,2008ApJ...680..568S,
2012AJ....143...17S}. The effect is less pronounced at longer 
wavelengths, as expected if the attenuation of the lines is 
due to dust extinction, and rules out that the blueshifts are 
due to electron scattering \citep{2012AJ....144..131Z} 
(Supplementary Information). 
The early epoch hydrogen lines have a Lorentzian half width 
at half maximum (HWHM) in the range 1,000--2,000 km s
$^{-1}$. The middle and right panels of Figure~1 show that 
the line profiles at the late epoch are narrower (HWHM 
$\sim800\pm100$\,km\,s$^{-1}$) and also exhibit blueshifts 
of the oxygen lines, which indicates that ejecta material is 
involved in the dust formation at this stage.
%

Figure~2 shows the temporal evolution of the inferred 
extinction, $A_{\lambda}$, 
as derived from the attenuation of 
emission lines in the early spectra. 
The extinction has been 
calculated from the ratios of the integrated line profiles at 
each epoch. We assume that the first epoch at 26 days past 
peak is nearly unextinguished and use it as a reference. The 
monotonic increase of the extinction as a function of time 
indicates continuous formation of dust. The extinction at 239 
days is $A_{V} \sim 0.6$ mag. Interestingly, the shape of 
the normalized extinction curve shows no substantial 
variation with time. Scaling and combining the data from the 
eight individual early epochs allows us to produce the first 
directly measured, robust extinction curve for a supernova. 
The extinction curve is shallow, with $R_{V} = A_{V}/E(B-V)
\approx 6.4$, and can be represented by a mix of 
grey-extinction dust grains ($A_{\lambda}$ = constant) and 
either standard Small Magellanic Cloud (SMC) or Milky Way (MW)
extinction grains \citep{2003ApJ...594..279G}. The extinction 
contribution of the grey dust is 40 \% in the $V$ band. We fit 
several dust models to the extinction curve using amorphous 
carbon dust characterized by a power-law grain size 
distribution \citep{1977ApJ...217..425M}
with slope $\alpha$, and minimum and maximum 
grain radii ($a_{\mathrm{min}}$ $<$ $a_{\mathrm{max}}$) in 
the interval [0.001, 5.0] $\mu$m. 

Figure~3 shows the resulting confidence interval for the two 
parameters $a_{\mathrm{max}}$ and $\alpha$ around the 
best fit values of $a_{\mathrm{min}}$ = 0.001 $\mu$m, 
$a_{\mathrm{max}}= 4.2$ $\mu$m and $\alpha=3.6$. It is 
evident that only size distributions extending to grain 
radii that are significantly larger than that of MW
interstellar medium \citep{2004ApJS..152..211Z, 
2012ApJ...744..129B} dust ($\gtrsim 0.25~\mu$m) can 
reproduce the supernova extinction curve (Figure~2).
The 2 $\sigma$ lower limit on the maximum grain size is 
$a_{\mathrm{max}}> 0.7~ \mu$m. We cannot perform a 
similar analysis of the late epoch because the intrinsic line 
profile at this epoch is unknown and likely highly affected by 
extinction \citep{2013ApJ...776....5M}. However, we note that 
the blueshift velocities change only marginally with 
wavelength (Extended Data Figure 6), suggestive of large grains 
also at this epoch.
%
 
Figure~4 illustrates the continuous build-up of dust as a 
function of time. 
The increasing attenuation of the lines is 
accompanied by increasing emission in the near-infrared 
(NIR) spectra, from a slight excess over a supernova 
blackbody fit at early times to total dominance at the late 
epoch. We fitted the spectra with black bodies which 
for the NIR excess yield a constant black-body radius of 
$(1.0 \pm 0.2) \times 10^{16}$ cm at the early epochs, and a 
temperature that declines from $\sim$ 2,300 K to 
$\sim$ 1,600 K from day 26 onwards. At the late epoch, we 
obtain a black-body radius of $(5.7 \pm 0.2) \times 10^{16}$ 
cm and a temperature of $\sim$ 1,100 K. The high 
temperatures detected at the early epochs suggest that the 
NIR excess is due to thermal emission from carbonaceous dust,
rather than silicate dust, which has a lower condensation 
temperature of $\sim 1,500$~K \citep{2011A&ARv..19...43G}.
The high temperatures rule out suggestions that the NIR emission is 
due to pre-existing dust or a dust echo \citep{2011AJ....142...45A} 
(Extended Data Figures 7, 8, Supplementary Information).
Fitting the NIR excess with a modified black body, assuming the 
grain composition found in our analysis of the extinction 
curve (Figure~3), gives a dust temperature similar to the 
black-body temperature, which is at all epochs 
(and considered dust compositions) larger than 1,000 K. 
The dust masses inferred from the extinction and NIR 
emission agree very well. 
The inferred amount of dust at the 
late epoch (868 days) is $\sim$ 2.5 $\times$ 10$^{-3}$~$\Msun$ 
if composed of carbon, but could be up to an 
order of magnitude larger for silicates 
(Methods).
Our results indicate accelerated dust formation after several 
hundred days. SN 2010jl will contain a dust mass of $\sim 0.5$~$\Msun$
similar to that observed in SN 1987A 
\citep{2011Sci...333.1258M, 2014ApJ...782L...2I}, 
by day $\sim 8000$, if the dust 
production continues to follow 
the trend depicted in Figure 4. 

The most obvious location for early dust formation is in a 
cool, dense shell behind the supernova shock 
\citep{2008ApJ...680..568S, 2009ApJ...691..650F},
which sweeps up material as it propagates through 
the dense circumstellar shell surrounding SN 2010jl 
\citep{2014ApJ...781...42O} (Supplementary Information). 
Dust formation in the ejecta is 
impossible at this stage because the temperature is too high.
The postshock gas cools and gets compressed to the low 
temperatures and high densities necessary for dust 
formation and gives rise to the observed intermediate width 
emission lines. By the time of our first observation 26 days 
past peak, the supernova blast wave encounters the dense 
circumstellar shell at a radius of $\sim$ 2.0 $\times$ 
10$^{16}$ cm for a blast wave velocity of $\sim 3.5 \times 
10^{4}$ km s$^{-1}$. As indicated by the blueshifts of the 
ejecta metal lines (Figure 1), the accelerated dust formation 
occurring at later times (Figure 4) and at larger radius is 
possibly facilitated by the bulk ejecta material, which
travels on average at a velocity of $\sim 7,500$ km s$^{-1}$ 
at early epochs (Extended Data Figure 4).

Our detection of large grains soon after the supernova 
explosion suggests a remarkably rapid and efficient 
mechanism for dust nucleation and growth. The underlying  
physics 
is poorly understood but may 
involve a two-stage process governed by early dust  
formation in a cool, dense shell, followed by 
accelerated dust formation involving ejecta material. 
For Type IIP supernovae, the growth of dust grains 
can be sustained up to 5 years past explosion 
\citep{2013ApJ...776..107S}. The dense CSM around 
Type IIn supernovae may provide conditions to facilitate 
dust growth beyond that. 
The process appears generic, in that other Type 	
IIn supernovae like SN 1995N, SN 1998S, SN 2005ip, and 
SN 2006jd exhibited similar observed NIR properties 
\citep{2004MNRAS.352..457P,2012ApJ...756..173S,
2012MNRAS.424.2659M,2013AJ....145..118V} and growing 
dust masses, consistent with the trend revealed here for SN 
2010jl (Figure~4). Moreover, it establishes a link between the 
early small dust masses inferred in supernovae 
\citep{2011A&ARv..19...43G, 2004MNRAS.352..457P,
2012ApJ...756..173S} and the large dust masses found in a 
few supernova remnants \citep{2011A&ARv..19...43G,
2011Sci...333.1258M, 2012ApJ...760...96G}. Large grains 
(0.1 $\le$ $a_{\mathrm{max}}$ $\lesssim$ 4.0 $\mu$m),
provide an effective way to counter destructive processes in 
the interstellar medium \citep{2011A&A...530A..44J}. Indeed, 
large grains from the interstellar medium have been detected 
in the Solar System \citep{1999ApJ...525..492F}.
Simulations indicate that grains larger than $\sim 0.1$ $\mu
$m will survive reverse shock interactions with only a low 
fraction being sputtered to smaller radii 
\citep{2010ApJ...715.1575S}. For a grain size distribution of 
$a_{\mathrm{min}}$= 0.001 $\mu$m, 
$a_{\mathrm{max}}= 4.2$ $\mu$m and  $\alpha=3.6$ 
(Figures~2 and 3), the mass fraction of grains above 
$0.1$ $\mu$m is $\sim$ 80 $\%$, i.e., the majority of 
the produced dust mass can be retained. 
%
%
\newpage
%
%
\section*{Methods summary}
We obtained optical and near-infrared medium-resolution 
spectroscopy with the ESO VLT/X-shooter instrument of the 
bright Type IIn supernova SN 2010jl at 9 epochs between 
2010 November 13.4 UT and 2011 June 15.0 UT. The continuum 
emission of the spectra was fit with a combination of black-body, 
modified black-body and host galaxy models, allowing us to 
quantify the temporal progression of the temperature and 
radius of the photosphere as well as the temperature and 
characteristics of the forming dust, which causes conspicuous 
excess near-infrared emission. We analysed the profiles of 
the most prominent hydrogen, helium and oxygen emission lines. 
From Lorentzian profile fits, which are good representations of the 
emission lines, we measured the blueshifts of the peaks and 
the half widths at half maximum of the lines,
and  derived the wavelength dependent 
attenuation properties of the forming dust at each epoch.
The uncertainties were obtained using Monte Carlo calculations 
by varying the Lorentzian profile parameters. 
We generated synthetic UBVRIJHK lighcurves 
and calculated the energy output of the supernova. This, together 
with calculated dust vaporization radii, temperatures of the dust 
grains at different distances from the supernova, and the radius 
evolution of the forward shock, were used to constrain the location 
of the forming dust. 
Different dust models, characterised by either single grain sizes or 
a power-law grain size distribution function 
and either amorphous carbon or silicates, were fitted to 
the extinction curves and the near-infrared excess emission. 
From these fits, we derived the temporal 
progression of the dust mass of the forming dust at each observed epoch.
%
\newpage

\bibliographystyle{naturemag}
\bibliography{sn2010jl}

%
\section*{Supplementary Information}
Is appended and available in the online version of the paper.

%
\section*{Acknowledgments}
We thank Lise Christensen and Teddy Frederiksen for advice 
on data reduction with the X-shooter pipeline and Maximilian 
Stritzinger and Rick Arendt for discussions. This investigation 
is based on observations made with ESO Telescopes at the 
La Silla Paranal Observatory under programme IDs 
084.C-0315(D) and 087.C-0456(A). C.G. was supported 
from the NASA Postdoctoral Program (NPP) and 
acknowledges funding provided by the Danish Agency for 
Science and Technology and Innovation. G.L. is supported 
by the Swedish Research Council through grant No.\ 
623-2011-7117. A.C.D.J. is supported by the Proyecto Basal 
PB06 (CATA), and partially supported by the Joint 
Committee ESO-Government Chile. The Dark Cosmology 
Centre is funded by the Danish National Research 
Foundation. 
%
%
\section*{Author contributions}
C.G. and J.H. conducted the observational campaign, 
reduced and analysed the data and wrote the manuscript. 
D.W. was the P.I. of the observing programs and assisted in 
writing the manuscript. 
E.D. performed calculations of vaporization radii and 
assisted in writing the manuscript. 
O.F. and G.L. assisted in data analysis.
J.R.M. helped with the interpretation of the spectra and line 
profiles.
D.M. and D.W. assisted with observations. 
A.C.D.J. conducted the observation of the epoch 2 
spectrum.
All authors were engaged in discussions and provided 
comments on the 
manuscript.  
%
%
\section*{Author information}
Reprints and permissions information is available at
www.nature.com/reprints. The authors declare no 
competing financial interests. Readers are welcome to 
comment on the online version of the paper. Correspondence
and requests for materials should be addressed to 
C.G. (cgall@phys.au.dk).
%
%
%
\newpage
\begin{center}
    \leavevmode
    \includegraphics[scale=0.9]{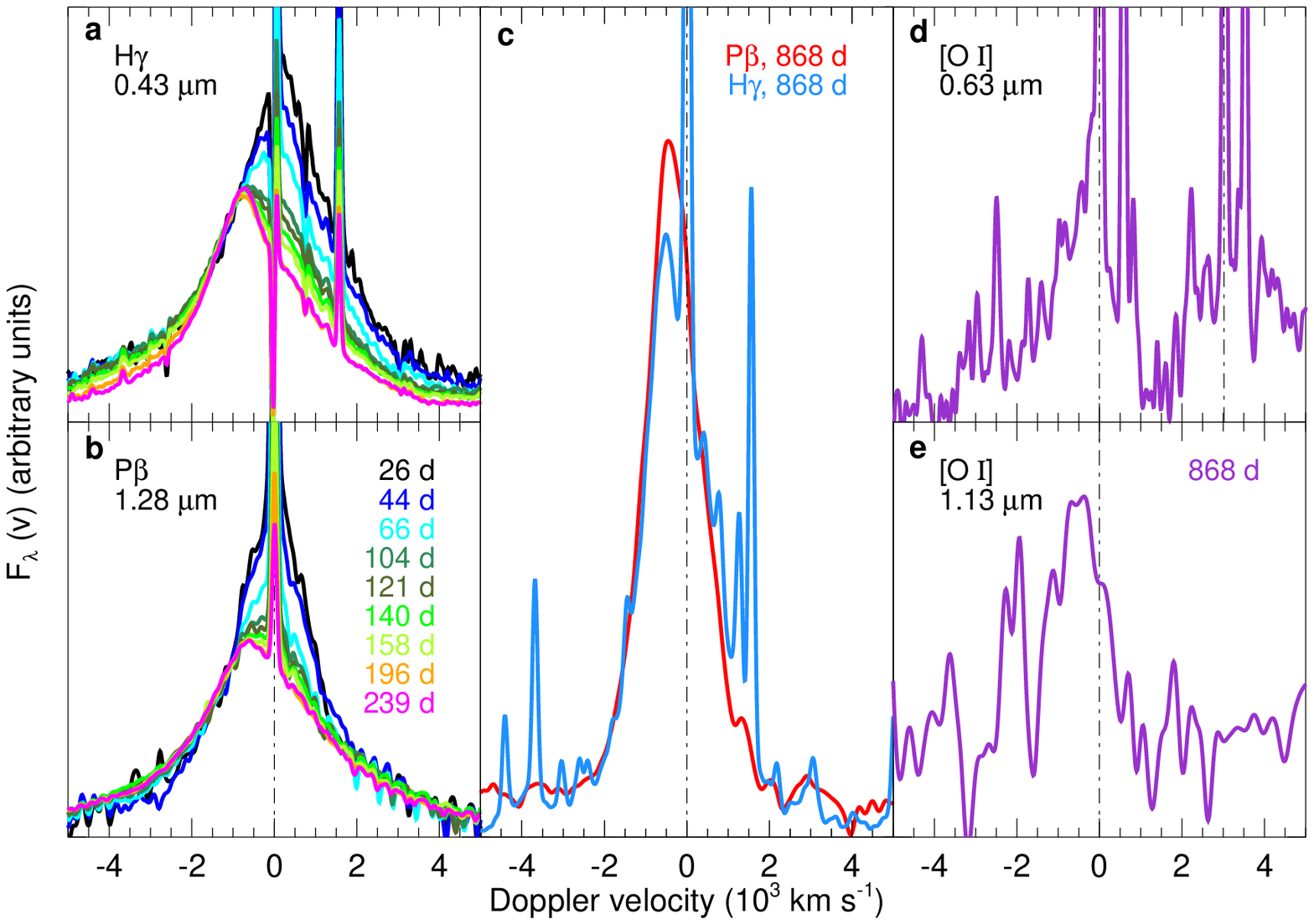}
\end{center}

{\noindent\bf Figure 1 $\mid$ 
 Evolution of the hydrogen and oxygen line profiles in the 
spectrum of SN 2010jl.} 
Line profiles for 
 {\bf a,} H$\gamma$ $\lambda$4340.472 and {\bf b,} P$\beta$ 
$\lambda $12818.072 for epochs from 26 to 239 days and
 {\bf c,} H$\gamma$ and P$\beta$ at 868 days.
{\bf d,} The 
[\ion{O}{i}] $\lambda\lambda$6300.304, 6363.776 doublet 
(zero velocity set at $\lambda$6300.304), and  {\bf e,} the [\ion{O}{i}] 
$\lambda$11297.68 line. 
The dashed-dotted lines in all panels denote zero velocity, at redshift 
$z$ = 0.01058, as determined from narrow emission lines in 
the spectrum.
%
%
\newpage
\begin{center}
    \leavevmode
 \includegraphics[scale=1.0]{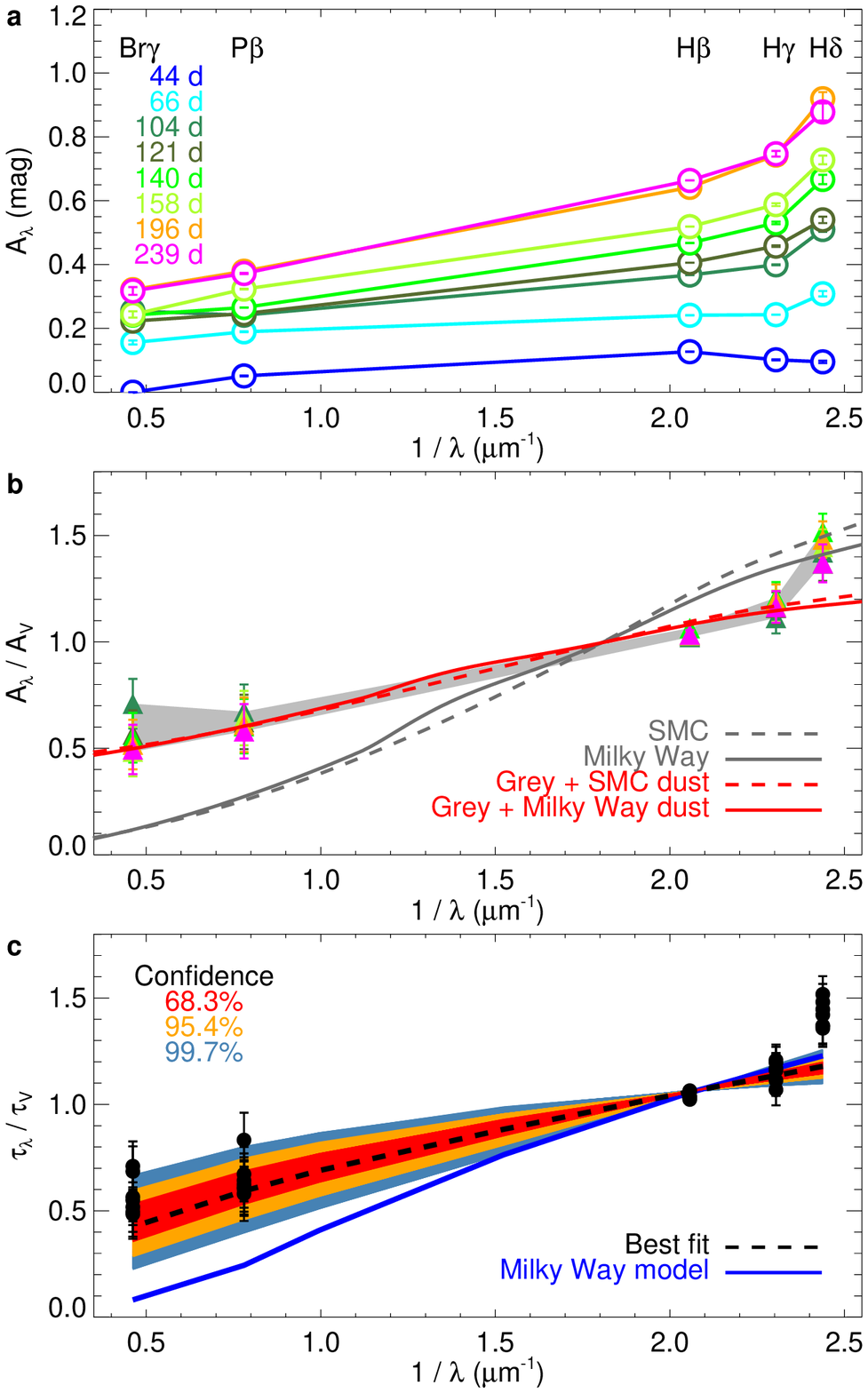}
\end{center}

{\noindent\bf Figure 2 $\mid$  Supernova dust extinction curves.}
{\bf a,} The evolution of the extinction, 
$A_{\mathrm{\lambda}}$, of the hydrogen lines 
(open circles with standard deviations; Methods).
The solid lines represent the (linearly interpolated) 
extinction curves.
{\bf b,} The grey-shaded area represents the range of 
extinction curves relative to $A_{V}$ 
(filled triangles with error bars). 
Grey curves are the SMC and MW extinction curves, 
while the red curves include a grey component (Methods).
{\bf c,}  Fits to the optical depth within the 1, 2 and 3 $\sigma$ 
(68.3, 95.4 and 99.7 $\%$) confidence interval (Methods). 
Dashed and solid curves are models with
`best fitting' and MW parameters, respectively. 
%
%
\newpage
\begin{center}
    \leavevmode
    \epsfxsize=\textwidth
    \includegraphics[scale=1.0]{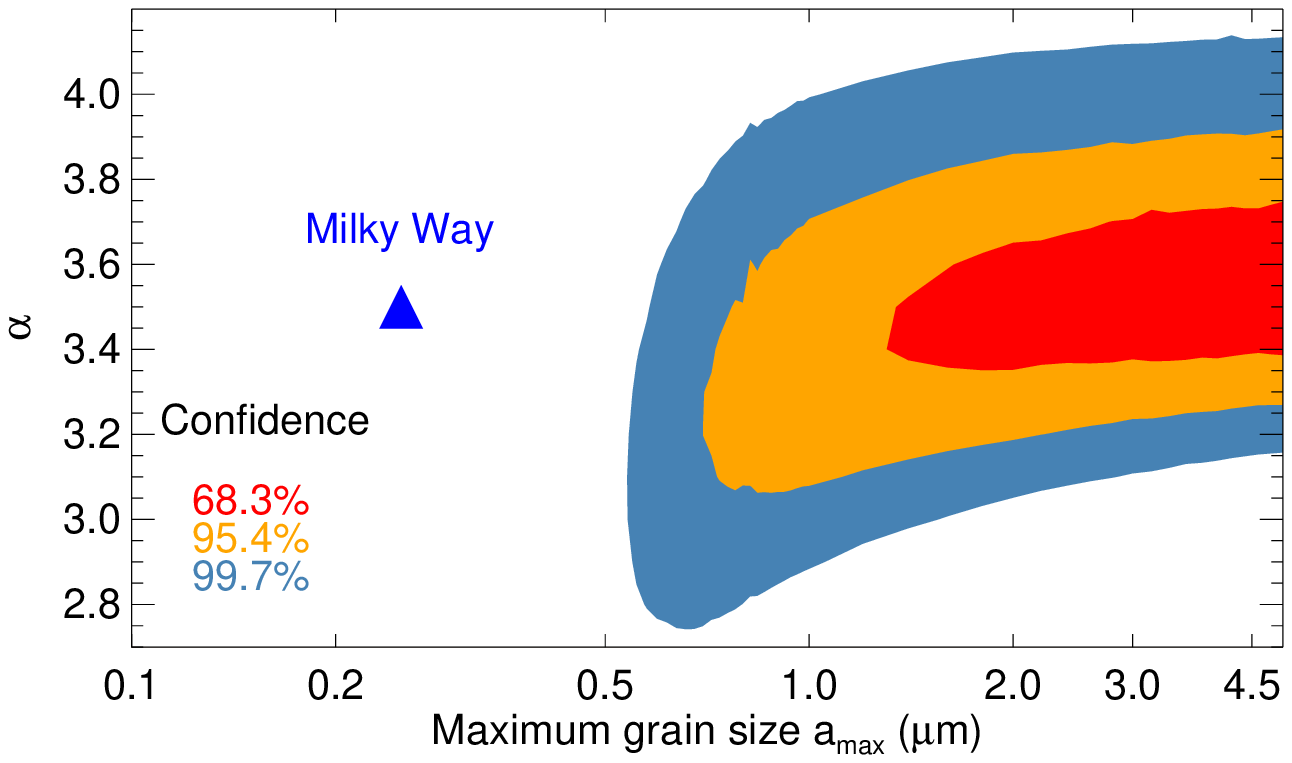}    
\end{center}

 {\noindent\bf Figure 3 $\mid$ 
 Maximum grain size and slope of the grain size 
distribution.}
Confidence contours, as constrained by the normalized 
optical depth $\tau(\lambda)$ (see Figure~2).
The most favorable power-law models lie within a 
parameter range for $\alpha$ between 
$\sim$ 3.4 and 3.7 and require large grains of 
$a_{\mathrm{max}}\gtrsim 1.3\,\mu$m (1$\sigma$). 
The confidence limits are as in Figure 2. 
Even at the $3\sigma$ confidence limit the 
maximum grain size is larger 
($a_{\mathrm{max}}\gtrsim0.5\,\mu$m) than MW
maximum grain sizes for a power law model 
($a_{\mathrm{max}}\approx 0.25\,\mu$m) 
\citep{1977ApJ...217..425M} or more sophisticated models
\citep{2004ApJS..152..211Z, 2012ApJ...744..129B}. 
%
%
\newpage
\begin{center}
    \leavevmode
       \includegraphics[scale=1.0]{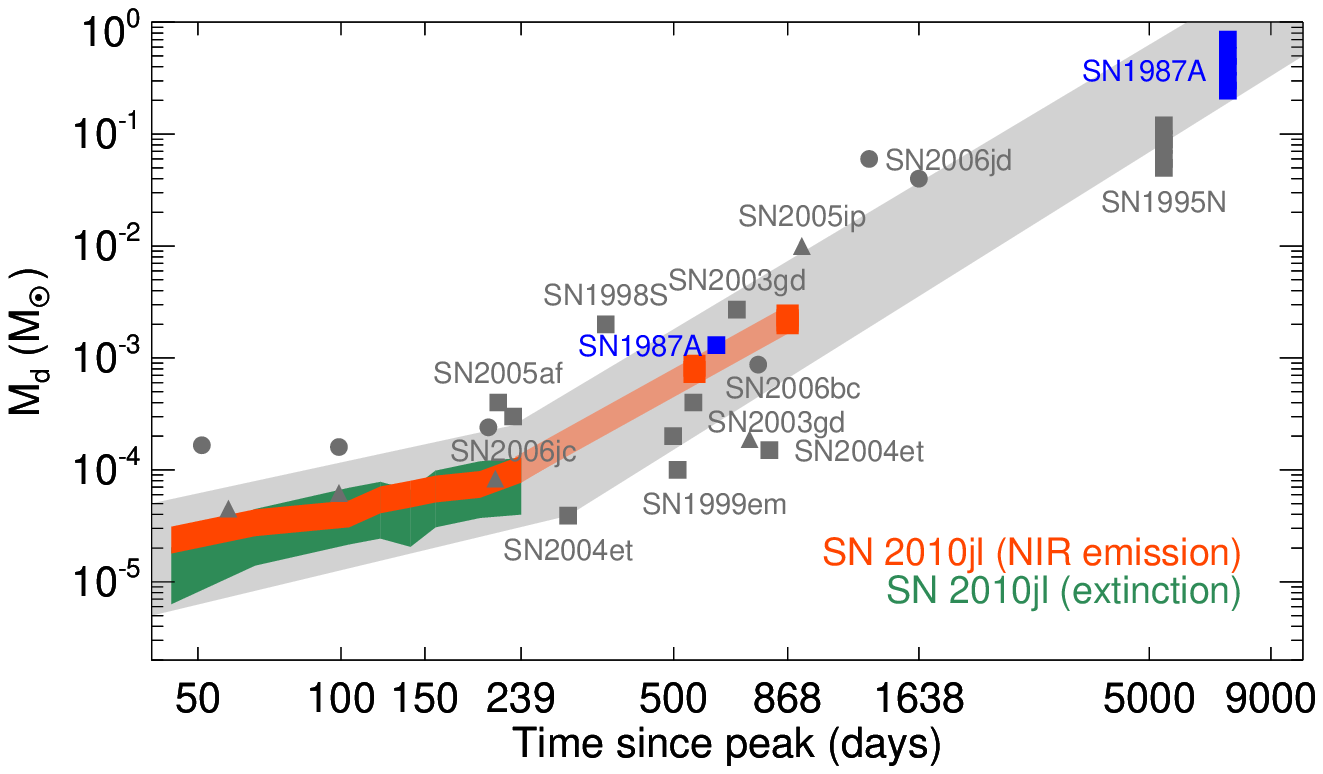}
\end{center}

 {\noindent\bf Figure 4 $\mid$ 
 Temporal evolution of the dust mass.}
Carbon dust masses and standard deviation 
derived from the extinction (green band) 
and the NIR emission 
(red bars and band; Methods) 
including a literature data point at 553 days 
\citep{2013ApJ...776....5M}. 
The light grey shaded area illustrates the 
evolution of the early  
($M_{\mathrm{d}} \propto t^{0.8}$ at $t<$ 250 days) 
and late 
($M_{\mathrm{d}} \propto t^{2.4}$ at $t>$ 250 days) 
stages of dust formation when SN 2010jl switches 
from circumstellar to ejecta dust formation. The grey 
and blue symbols correspond 
to literature data 	
for SN 2005ip (triangles), SN 2006jd (dots), 
and other supernovae (bars)	
\citep{2011Sci...333.1258M,2011A&ARv..19...43G,
2014ApJ...782L...2I, 2012ApJ...744...26O,2012ApJ...753..109G,
2012ApJ...756..173S,2013AJ....145..118V}. 
The length of the symbols for SN 1995N and 
SN1987A correspond to the quoted dust mass range.
For other supernova the standard deviation is either
smaller than the size of the symbols or have not been 
reported. 
%
%
\newpage
\begin{methods}

We observed the Type IIn SN 2010jl in UGC 5891A at 9 epochs 
between 2010 November 13.4 UT and 2011 June 15.0 UT, 
following its discovery on 2010 November 3.5 UT 
\citemeth{2010CBET.2536....1B}. The SN was first detected 
from the All Sky Automated Survey North on 2010 October 9.6 
UT and peaked on 2010 October 18.6 UT 
\citemeth{2011ApJ...730...34S}. The time of explosion is unknown, 
but we assume a time of rise to peak of $\approx$ 40 days.
We adopt a luminosity distance of $D=45.7$ Mpc to the SN, 
based on our measured redshift of $z = 0.01058$.

The observations were obtained with the X-shooter echelle 
spectrograph \citemeth{2006SPIE.6269E..98D,
2011A&A...536A.105V} mounted at the Cassegrain focus of 
the {\it Kueyen} unit of the Very Large Telescope (VLT) at the 
European Southern Observatory (ESO) on Cerro Paranal, 
Chile. The X-shooter instrument allows for simultaneous 
spectroscopic observations in three different arms, the 
ultra-violet and blue (UVB), visual (VIS) and near-infrared (NIR) 
wavebands, covering the continuous wavelength range of 
0.3--2.5 $\mu$m. The observations were performed at 
parallactic angle, with nodding between exposures along the 
11$'' $ slit (see Extended Data Table~1 for details). The spectra 
were obtained under the following conditions: clear sky and in 
some cases thin cirrus, average seeing of $\sim$ 0.8$'' $, and 
range in airmass of $\sim$ 1.2--2.0.  For the majority of the 
observations we used slit widths of 1.0$'' $ (UVB), and 0.9$'' $ 
(VIS and NIR) giving resolving powers of 5,100 (UVB), 8,800 
(VIS) and 5,300 (NIR), except for the second epoch on 2010 
December 1.4 UT where, due to mediocre seeing conditions of 
$\sim$ 1.7$'' $, we used wider slit widths of 1.6$'' $ (UVB), and 
1.5$'' $ (VIS and NIR), leading to reduced resolving powers of  
3,300 (UVB), 5,400 (VIS) and 3,500 (NIR). For all epochs, 
observations of spectrophotometric standards were performed 
using a slit width of 5.0$'' $.   

We used versions 1.5.0 and 2.2.0 of the X-shooter pipeline 
\citemeth{2010SPIE.7737E..56M} in physical mode to reduce the 
SN and the standard star spectra to two-dimensional bias-
subtracted, flat-field corrected, order rectified and wavelength 
calibrated spectra in counts. To obtain one-dimensional spectra 
the two-dimensional spectra from the pipeline were optimally 
extracted \citemeth{1986PASP...98..609H}. Furthermore the spectra 
were slit-loss corrected, flux calibrated and corrected for 
heliocentric velocities. Additionally, telluric 
corrections were applied. 
All calibration and correction procedures after the basic 
pipeline reduction were performed using custom IDL programs. 
The spectra were corrected for a Galactic extinction along the 
line of sight to the SN of $E(B-V)$ of 0.027 mag 
\citemeth{1998ApJ...500..525S}.

%
\section*{Two-temperature black-body fits}
The progressive evolution of the supernova spectra is shown in 
Extended Data Figure~1. 
We fit the continuum emission of the early epochs (26--239 
days) with a combination of two black-body functions. 
The first black body represents the supernova photosphere for 
which we infer a temperature $T_{\mathrm{SN}} \approx 7,300$ K 
and a photospheric radius decreasing from
$R_{\mathrm{SN}}$ $\sim$ 3.2 $\times$ 10$^{15}$ cm at 26 days 
to $R_{\mathrm{SN}}$ $\sim$ 2.4 $\times$ 10$^{15}$ cm at 239 
days. The second black-body function accounts for the NIR 
excess noticeable in the spectra, which we attribute to dust
emission.

To properly fit the hot dust emission we therefore
use a modified black-body function. The fit to 
the spectra is computed as 
\begin{equation}
F_{\mathrm{\nu}} (\nu)  =  B_{\mathrm{\nu}}(\nu, T_{\mathrm{SN}}) \, 
R_{\mathrm{SN}}^{2} / D^{2}  + N_{\mathrm{d}} /  D^{2}  
\int_{a_{\mathrm{min}}}^{a_{\mathrm{max}}} \, f(a) \, m(a)  \,  
\kappa_{\mathrm{abs}}(\nu, a) \, B_{\mathrm{\nu}}(\nu, T_{\mathrm{hot}})  
\, \ud a, 
\end{equation}
where $B_{\mathrm{\nu}}(\nu, T)$ is the Planck function at 
temperature $T = T_{\mathrm{SN}}$ for the supernova and 
$T = T_{\mathrm{hot}}$ for the dust, $R_{\mathrm{SN}}$ is the 
radius of the supernova photosphere, $D$ is the luminosity 
distance to the supernova, 
$N_{\mathrm{d}}$ is the total number of dust particles, 
$m_{\mathrm{d}}(a)$ =  $(4 \pi/3) \rho a^3$ is the mass of a 
single dust grain of radius $a$ and $\kappa_{\mathrm{abs}}(\nu)$ is the dust 
mass absorption coefficient for an assumed 
dust composition, i.e., amorphous carbon  \citemeth{1991ApJ...377..526R}
and silicates \citemeth{2001ApJ...554..778L}. The
mass density is $\rho$ = 1.8 g cm$^{-3}$ for amorphous carbon and 
$\rho$ = 3.3 g cm$^{-3}$ for silicates. 
We use a power-law grain size distribution function 
$f(a)$ $\propto$ $a^{-\alpha}$, which is normalized to unity in the 
interval $[a_{\mathrm{min}}$, $a_{\mathrm{max}}]$ as 
$\int_{a_{\mathrm{min}}}^{a_{\mathrm{max}}}  f(a) \, \ud a = 1$.

Extended Data Figure~2 depicts supernova spectra obtained at 
44 days, 196 days and at a late epoch of 868 days. 
The adopted grain size distribution assumes the parameters
$\alpha, a_{\mathrm{min}}$ and $a_{\mathrm{max}}$ from the
`best fitting' amorphous carbon model obtained 
from the extinction curves (Figures 2 and 3). 
To fit the spectrum at 868 days we exchanged the SN black-body 
with a power law for the host galaxy continuum emission as 
$C_{\mathrm{norm}} \, \nu^{-1.24}$,  
where $C_{\mathrm{norm}}$ is a normalization constant and 
the power law exponent resulted from the fit. 
Additionally, for this epoch we explored two dust compositions, 
i.e., amorphous carbon and silicates, and models with 
single grain sizes $a$ between 0.001 and 5.0 $\mu$m, 
as well as grain size distribution models varying 
 $\alpha$ between 2.0 and 4.5.  
We found that 
(1) amorphous carbon single grain size models as well as 
grain size distribution models prefer large grains 
(1--5 $\mu$m),
(2) the quality of the fits of silicate models is fairly insensitive to 
the size of the grains and can accommodate small grains,
(3) we are unable to produce models with temperatures less 
than $\sim$ 1,000 K,
(4) the inferred dust masses for silicate grains are typically 
up to an order of magnitude higher than for amorphous carbon. 
All spectra are well fit by 
a supernova temperature $T_{\mathrm{SN}}$ $\approx$ 7,300~K and a 
dust temperature, $T_{\mathrm{hot}}$, which decreases from 
$\approx$ 2,300 K to 1,600 K during the first 239 days, 
down to $\approx$ 1,100~K at 868 days.

Extended Data Figure 1 also shows {\em Spitzer}/IRAC 3.6 $\mu$m and 
4.5 $\mu$m observations \citemeth{2011AJ....142...45A}. We can fit 
the 3.6 $\mu$m data point with the same modified black-body model 
as used for the other epochs (grey dotted curve in Extended 
Data Figure~1). 
%
%
\section*{Analysis of line profiles}

The spectra exhibit a richness of emission lines on top of the 
continuum, featuring in particular hydrogen and helium lines 
which are:
H$\delta$ $\lambda$4101.734, 
H$\gamma$ $\lambda$4340.472, 
H$\beta$ $\lambda$4861.35, 
\ion{He}{i} $\lambda$5875.621,
H$\alpha$ $\lambda$6562.79,  
\ion{He}{i} $\lambda$7065.2578, 
P$\delta$ $\lambda$10049.8,
\ion{He}{i} $\lambda$10830.199, 
P$\gamma$ $\lambda$10938.17,
P$\beta$ $\lambda$12818.072 and 
Br$\gamma$ $\lambda$21655.268. 
The lines have a narrow ($\sim$ 100 km s$^{-1}$) velocity 
component on top of an intermediate width velocity component. 
For H$\delta$, H$\gamma$, H$\beta$, H$\alpha$,  
\ion{He}{i} $\lambda$5875.621 and 
\ion{He}{i} $\lambda$10830.199 the narrow lines exhibit a 
characteristic P~Cygni profile (i.e., a blueshifted absorption and 
redshifted emission component). 

Only a subset of the hydrogen lines are suitable for quantitative 
extinction studies. We require that the lines exhibit a clear single 
peaked intermediate velocity component across all epochs which can be 
well represented by a Lorentzian profile, not necessarily centred  
at the zero velocity (Extended Data Figure~3a). 
None of the  \ion{He}{i} lines are suitable for extinction
studies because they show conspicuous bumps in the wings.
Moreover, the wings significantly broaden with time 
(Extended Data Figure~3b). 
The \ion{He}{i} $\lambda$10830.199 line is blended 
with the P$\gamma$ $\lambda$10938.17 line, ruling out both 
lines for our studies. The P$\delta$ $\lambda$10049.8 line is 
located at the cross-over between the X-shooter VIS and NIR 
arms giving rise to unreliable flux calibration and background 
subtraction.

Some lines show the presence of large velocities. 
Extended Data Figure~4 shows H$\beta$ P~Cygni profiles
featuring velocities up to $\sim$ 20,000 km s$^{-1}$ 
which arise from the fast expanding thin outer layers of 
the supernova ejecta. The bulk expansion velocity of the 
supernova ejecta (corresponding to the minimum of the 
P~Cygni profile) is around 7,500 km s$^{-1}$.
Other lines, e.g., H$\alpha$, are 
characterized by an underlying broad velocity component. As shown in 
Extended Data Figure~5a, we can fit the 26 day H$\alpha$ line with a 
combination of a broad Gaussian with a full with of half maximum of 
$\sim$ 5,000 km s $^{-1}$ and an intermediate Lorentzian with 
HWHM $\sim$ 860 km s$^{-1}$ centred at zero velocity. 
While the H$\alpha$ line is 
often used to demonstrate the effect of 
dust attenuation of the red wing \citemeth{2013arXiv1312.6617F},
it is discarded for our study because of the 
progressive broadening of the wings (Extended Data Figure~3b),
which prevents a straightforward quantitative analysis.

At 868 days the emission lines no longer exhibit a broad 
velocity component. The intermediate velocity components 
of the hydrogen emission lines feature velocities up to 
$\sim$ 2,000--3,000  km s$^{-1}$ similar to the 
oxygen [\ion{O}{i}] $\lambda$6300.304 and [\ion{O}{i}] 
$\lambda$11297.68 lines (Figure~1). The lines are not well
represented by Lorentzian profiles (Figure 1, middle panel). 
Consequently, the late epoch is not considered for our quantitative 
extinction studies.  

From single Lorentzians fits to 
H$\delta$, H$\gamma$, H$\beta$, P$\beta$ and Br$\gamma$, 
we estimated the Lorentzian HWHM 
of the intermediate velocity components of 	
these lines (about 1,500 $\pm$ 200 km s$^{-1}$).
Extended Data Figure~5b shows that the hydrogen lines 
(e.g., H$\beta$) exhibit deviations from symmetry, despite 
being adequately represented by Lorentzian profiles for 
our purposes (Extended Data Figure~3a). We also measured 
the blueshifts of the peaks (Extended Data Figure 6).

To obtain the hydrogen line profiles (Figure 1), the spectrum 
from each epoch was continuum subtracted and scaled to the 
first epoch. The scaling was set by the velocity at which the blue 
side of the line changed from being extinguished to being 
unextinguished (between $-1,200$ to $-1,000$ km s$^{-1}$).
This ensures that we measure only the extinguished parts 
of the lines. The blue unextinguished wings from all epochs 
coincide. At the late epoch (868 days), H$\gamma$ was scaled 
to P$\beta$ at a velocity of $-800$ km s$^{-1}$. 

\section*{Extinction measurements}

Attributing the red depressions to dust, we calculated the 
extinction (Figure 2a) from the fitted Lorentzian profiles as
$A_{\lambda} = - 2.5~\log(I(\lambda,t) / I_{\mathrm{ref}}(\lambda))$, 
where $I(\lambda,t)$ is the line profile integrated over a velocity 
range extending from the scaling velocity up to
4,000 km s$^{-1}$ and $I_{\mathrm{ref}}(\lambda)$ 
is the integrated line profile from the first epoch which was
taken as a reference. We obtained the error bars of $A_{\lambda}$ 
(standard deviations) using Monte Carlo calculations by varying 
the fit parameters of the Lorentzian line profiles within their 
uncertainties. The error bars reflect the signal to noise ratio of 
the lines and the extent to which they are well represented by 
Lorentzians. 
From measurements of $A_{V}$ ($\lambda_{V} = 5,505$ \AA) 
and $ E(B-V)$ in Figure 2, we directly infer 
$R_{V}$ = $A_{V}/E(B-V)$ $\approx 6.4$. The  wavelength 
dependent optical depth is fit with
(1) a phenomenological model based 
on grey dust plus either a SMC or MW dust, 
$A_{\mathrm{SMC, MW}}(\lambda)$, as 
$A_{\lambda} = A_{\mathrm{grey}} + A_{\mathrm{SMC, MW}}(\lambda)$
(Figure 2b) and 
(2) a single dust model, i.e., only carbon dust, 
which for a shell is,
\begin{equation}
\tau({\lambda}) = \frac{N_{\mathrm{d}}}{4 \, \pi \, R^{2}}  \int_{a_{\mathrm{min}}}^{a_{\mathrm{max}}}  f(a) \, m_{\mathrm{d}}(a) \,  \kappa_{\mathrm{ext}}(\lambda, a) \, \ud a,
\end{equation}
where
$R$ is the distance of the cool dense shell (CDS) from the SN 
and $\kappa_{\mathrm{ext}}(\lambda, a)$ is the mass absorption 
coefficient, i.e., for amorphous 
carbon \citemeth{1991ApJ...377..526R}. 
We calculate a grid of models varying the slope $\alpha$ between
0.5 and 4.5, and
the lower and upper limits of the grain size distribution, 
$a_{\mathrm{min}}$ and $a_{\mathrm{max}}$,
between 0.001 and 5.0 $\mu$m
($a_{\mathrm{min}} < a_{\mathrm{max}}$).
The dispersion of the normalized data (between day 66--239) is 
added to the error (Figure 2c). 
We used the chi-square ($\chi^{2}$) minimization method
to determine the best fitting parameters $\alpha$ and $a_{\mathrm{max}}$ 
of the grain size distribution function, for fixed
$a_{\mathrm{min}} = 0.001\,\mu$m (Figure 3). 
The $\chi^{2}$ value for a desired confidence limit $p$ is calculated as 
$\chi^{2} = \chi_{\mathrm{min}}^{2} + \Delta \chi^{2}(p)$, 
where $\chi_{\mathrm{min}}^{2}$ is the global minimum 
$\chi_{\mathrm{min}}^{2}$ value of all models. 
The best fitting model is characterized by
$\alpha = 3.6, a_{\mathrm{min}} = 0.001\,\mu$m and 
$a_{\mathrm{max}} = 4.2\,\mu$m. 
However, our models cannot account for the upturn 
towards H$\delta$ (Figure 2b), which we attribute to a systematic 
effect caused by intrinsic line changes rather than to 
small grains.
We note that the considered 
grain radius is truncated at $5\,\mu$m beyond which the size 
parameter, $x = 2 \pi a/ \lambda$, becomes prohibitively large 
making mass absorption coefficient calculations difficult.
%

\section*{Lightcurves}
In Extended Data Figure~7a we show synthetic UVBRI optical 
and JHK NIR lightcurves generated from our X-shooter spectra
compared to broad band photometry from the literature
\citemeth{2013ApJ...776....5M} (we have added 1.4 mag to the published 
U band magnitudes, which happens to be twice the U band AB offset).
It is evident that there is good agreement giving 
credence to the flux calibration of our spectra.
The energy input (Extended Data Figure~7b)
from $^{56}$Co was normalized to the total observed luminosity on 
day $\sim 26$, and the $^{44}$Ti contribution was calculated 
assuming a relative Co/Ti yield (by number) of 
$3\times 10^{-4}$ \citemeth{2002ApJ...576..323R}.
The UVO and NIR luminosities (Extended Data Figure~7a) are
derived from the black body fits to our X-shooter spectra.
A power-law approximation to the UVO luminosity shows that it
decays as a $t^{-0.4}$ power-law.

\section*{Dust heating and vaporization}	

A dust grain of radius $a$ located at a distance $R$ from the SN 
will attain an equilibrium temperature, $T_{\mathrm{d}}$,
determined by the balance between the rate it is heated by the 
SN and its cooling rate by NIR emission. The equation 
describing this balance is given by
\begin{equation}
 \int_0^{\infty}\ \pi\, a^2\, Q_{\mathrm{abs}}(\nu,a)\, \left({L_{\nu}(\nu)\over 4 \pi R^2}\right)\, d\nu =  \int_0^{\infty}\ 4 \, \pi a^2\, Q_{\mathrm{abs}}(\nu,a)\, \pi\, B_{\nu}(\nu, T_{\mathrm{d}})\, d\nu.
\end{equation}
The SN will vaporize a grain when its temperature exceeds the 
vaporization temperature. We take 
$T_{\mathrm{vap,SIL}}$ = 1,500~K for silicates. 
Due to the large uncertainty in  
$T_{\mathrm{vap,AC}}$ for amorphous carbon grains we adopt a 
temperature range of 2,000--3,000~K. 

The SN light is preceded by a short ($\Delta t \lesssim 1$~d) 
burst of radiation as the shock, resulting from the core collapse, 
breaks out of the stellar surface \citemeth{1983ApJ...274..175D}. 
From direct observations  \citemeth{2008Natur.453..469S, 2009ApJ...692L..84M} 
of such a shock breakout and models to fit early UV optical lightcurves  
\citemeth{2000ApJ...532.1132B, 2013MNRAS.429.3181T} a shock 
breakout burst typically lasts of order 100--1000 seconds with 
peak luminosities of around $10^{12}$~$\Lsun$ and effective 
temperatures of a few times 10$^5$~K after which the 
luminosities decrease in less than one day by over a few orders 
of magnitudes. 
A short burst of radiation characterized by a $10^5$~K black body and a 
luminosity of $10^{11}$~$\Lsun$, similar to that inferred for 
Cas~A \citemeth{2008ApJ...685..976D}, will vaporize any 
pre-existing dust in the circumstellar material (CSM), creating a 
dust-free `cavity' of radius $R_{\mathrm{cav}}$. 
Extended Data Figure~8a shows that, independently of the grain 
species, small grains are vaporized out to larger distances from
the SN than large grains. 
For silicate grains, $R_{\mathrm{cav}}$ is about twice 
as large as for carbon grains. As a consequence of shock
breakout, no dust grains can exist out to $R_{\mathrm{cav}}$ of 
about 10$^{17-18}$ cm. 

Any dust that may subsequently form within the radius 
$R_{\mathrm{cav}}$ will be subjected to the SN light, 
characterized by a 7,300~K black body and a 
luminosity of $\sim 5 \times 10^9$~L$_{\odot}$.
Extended Data Figure~8b shows the vaporization radii, 
$R_{\mathrm{vap}}$, for the observed SN luminosity 
at the first epoch (26 days past peak). 
Independently of grain size, the $R_{\mathrm{vap}}$ 
for silicates is significantly larger than 
$R_{\mathrm{vap}}$ for amorphous carbon 
at any assumed $T_{\mathrm{vap,AC}}$ and 
CDS radius, which is about 2 $\times$ 10$^{16}$ cm 
at this epoch
(Supplementary Information, Extended Data Figure 9b).
Dust grains of sizes around 0.05--0.1 $\mu$m have the 
largest vaporization radii for either dust species. 
It is evident that only amorphous carbon grains 
can survive the radiation from the underlying SN
at the location of the CDS.
Amorphous carbon grains with grain sizes $\gtrsim 0.25~\mu$m have 
temperatures ($\lesssim$ 2,200~K), consistent with the hot dust 
temperatures inferred from the modified black-body fits to the 
NIR emission (Extended Data Figure~8c). 
These grain radii are consistent with those inferred 
from our extinction measurements.  
The small carbon grains ($\lessapprox 0.25~\mu$m), 
which are required to explain the observed UV extinction 
(see Figure~2), have higher temperatures (2,200--2,700~K) but
do not contribute strongly to the NIR emission.
Silicate grains cannot exist at the location of the CDS. 

%
\section*{Dust mass estimates} 
The dust mass is derived from either the extinction or the NIR 
emission as 
\begin{equation}
M_{\mathrm{d}}  = N_{\mathrm{d}} \,  \int_{a_{\mathrm{min}}}^{a_{\mathrm{max}}} \, f(a) \, m_{\mathrm{d}}(a) \, \ud a .
\end{equation}
The total number of dust particles $N_{\mathrm{d}}$ 
is obtained from the fits using either Equations (1) or (2). 

In Figure~4 we display the evolution of the carbon dust masses,
which are derived from
(1) the extinction and its standard 
deviation obtained for the best fitting grain size distribution
(Figures~2 and 3) at $R_{\mathrm{CDS}} = 2.0\times10^{16}$\,cm and
(2) the NIR emission, varying $\alpha$ between 3.5 and 3.7.
A power law fit, $M_{\mathrm{d}}$ $\varpropto$  $t^{\beta}$, 
to the dust mass evolution at early and late phases shows
a slow increase ($\beta$ = 0.8) at early times and accelerated build 
up of the dust mass ($\beta$ = 2.4) after 239 days. 
The estimated carbon dust mass of 
$\approx 2 \times 10^{-3}~\Msun$ at the late epoch is most likely 
a lower limit (Supplementary Information).

Extended Data Figure~9a visualises the sensitivity of the 
inferred extinction dust mass to $a_{\mathrm{max}}$ and $\alpha$. 
Carbon dust masses at 239 days past peak are calculated and displayed 
for parameters within the 3$\sigma$ confidence interval (see 
Figure~3) and for fixed $a_{\mathrm{min}}$ = 0.001 $\mu$m. 
Large grains exhibit a stronger dependency on $\alpha$, with larger 
dust masses being reached for small $\alpha$ 
(i.e., favouring large grains). For small grains the dust mass is 
almost independent of  $\alpha$. For large $\alpha$ the dust mass 
remains independent of $a_{\mathrm{max}}$ for large grains 
whereas for small $\alpha$, the dust mass increases steeply 
with increasing $a_{\mathrm{max}}$. 

Requiring that the extinction and emission dust 
mass originate from the CDS, the allowed location of the CDS is constrained by  
$R_{\mathrm{vap}}$ $\lesssim$ 
$R_{\mathrm{CDS}}$ $\lesssim$ $R_{\mathrm{shock}}$ 
(Extended Data Figure~9b). 
The location of the forward shock, $R_{\mathrm{shock}}$,
at day 239 is estimated assuming a velocity of      
$3.5 \times 10^{4}$ km s$^{-1}$ until 26 days past peak and     
3,000 km s$^{-1}$ for the subsequent 213 days.
%

%
\newpage
\begin{center}
\leavevmode
  \includegraphics[scale=1.0]{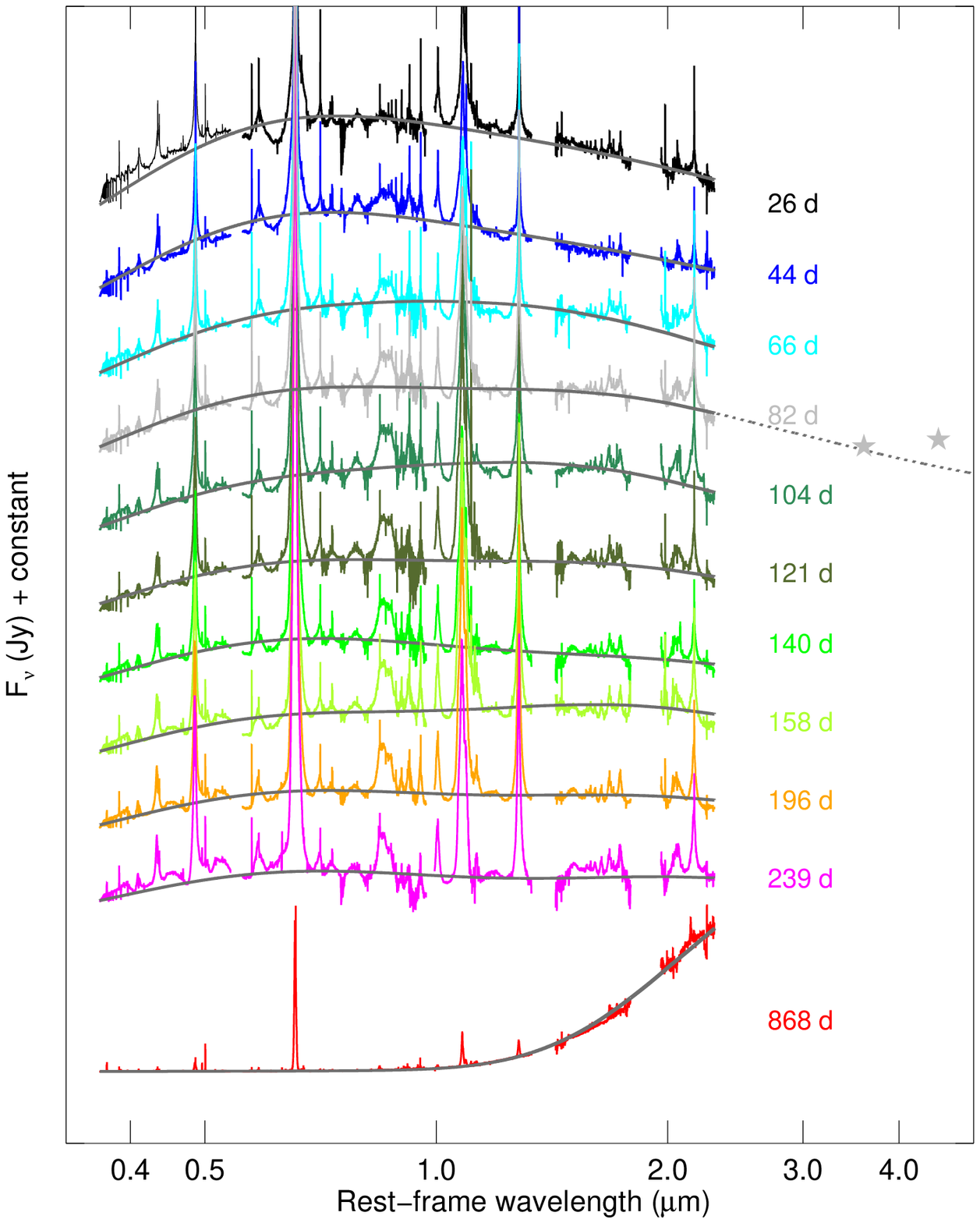}
\end{center}
\vspace{-1.5 cm}
 {\noindent\bf Extended Data Figure 1 $\mid$ 
 Time sequence of the supernova spectra.}  
Spectra from ten epochs between 
$t$ = 26 and 868 days past peak. The spectra are offset by an 
arbitrary constant. The atmospheric telluric bands at 1.33--1.43 
$\mu$m and 1.79--1.96 $\mu$m have been excluded as well 
as the dichroic gaps between the X-shooter instrument arms at 
0.54--0.56 $\mu$m and 0.97--0.995 $\mu$m.  
The light grey spectrum is an interpolated 
spectrum at the epoch of observations of the IRAC 3.6 $\mu$m 
and 4.5 $\mu$m data (grey stars) \citemeth{2011AJ....142...45A}. 
The solid grey curves are fits to the spectra, composed of multiple 
distinct black-body functions.
%
%
\newpage
\begin{center}
\leavevmode
    \includegraphics[scale=0.88]{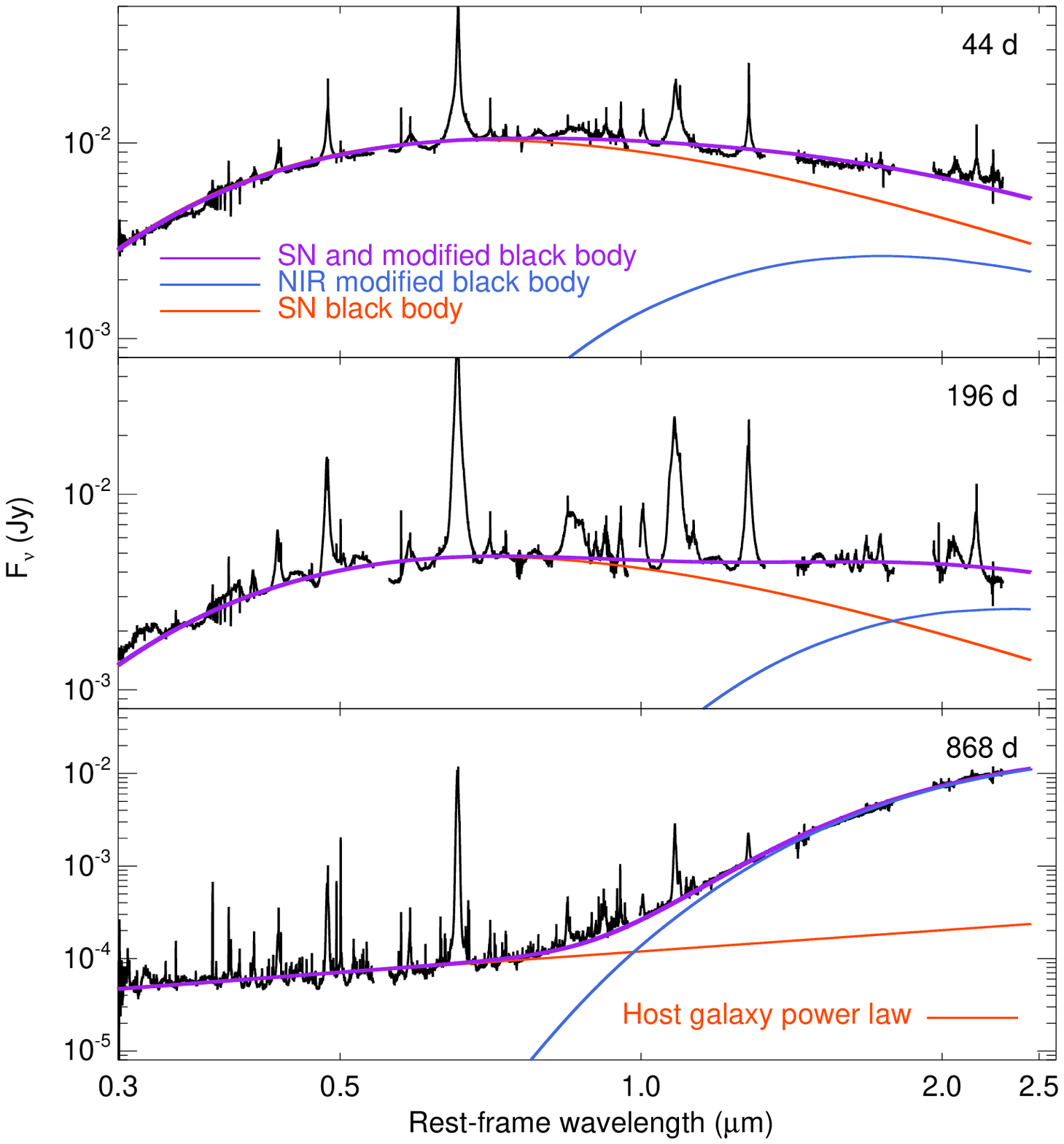}
\end{center}
\vspace{-2.0 cm}
{\noindent\bf Extended Data Figure 2 $\mid$  
NIR excess dust emission in supernova spectra at three different epochs.}  
The spectral shape of the supernova shows little evolution for the 
early epochs (44 and 196 days past peak). 
The late epoch at 868 days exhibits strong 
NIR emission while the supernova continuum has faded. 
The  atmospheric 
telluric bands at 1.33--1.43 $\mu$m and 1.79--1.96 $\mu$m as 
well as the dichroic gaps of the X-shooter instrument arms at 
0.54--0.56 $\mu$m and 0.97--0.995 $\mu$m have been 
excluded. 
%
%
%
\newpage
\begin{center}
\leavevmode
\includegraphics[scale=0.75]{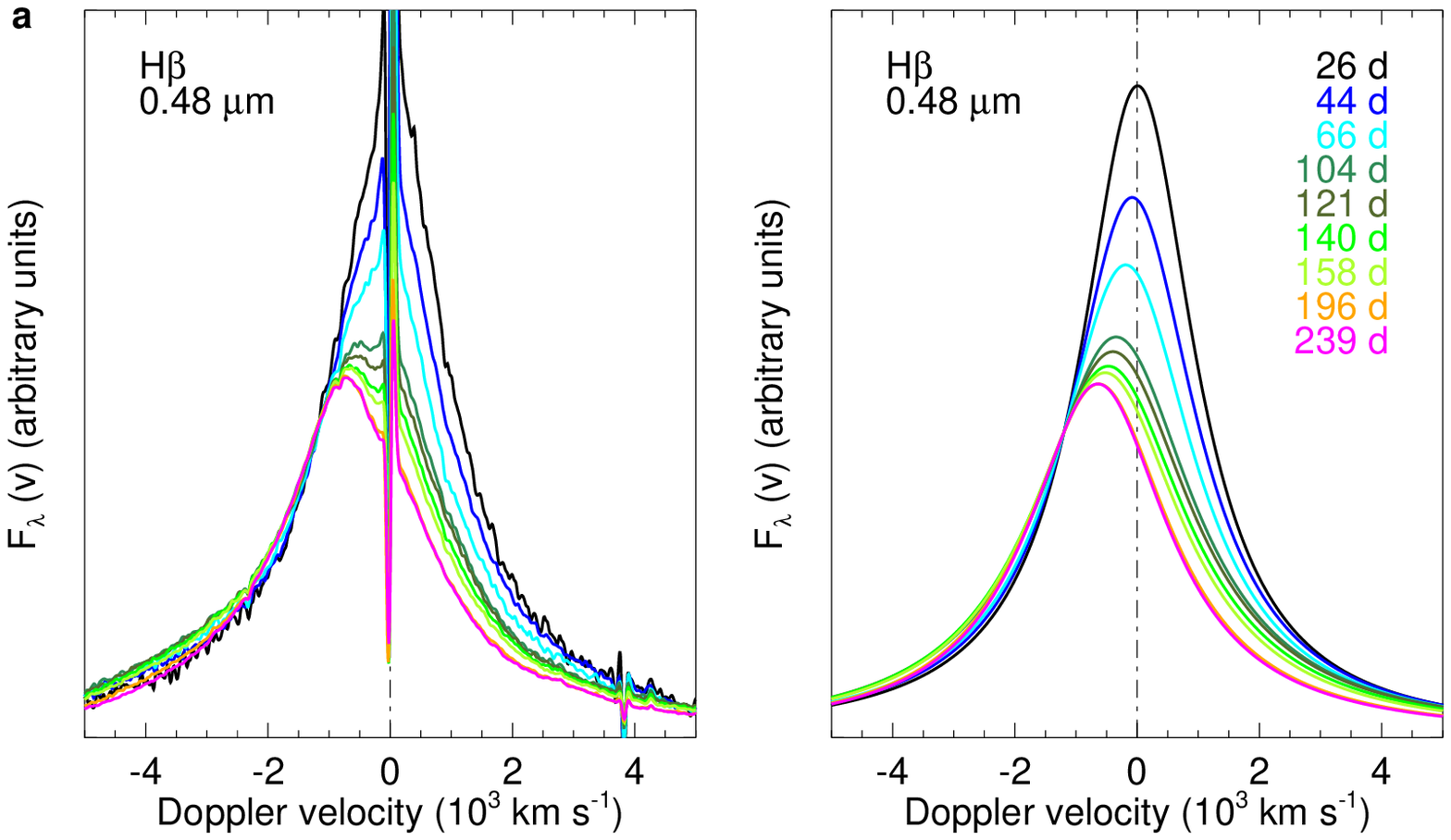}   
\includegraphics[scale=0.75]{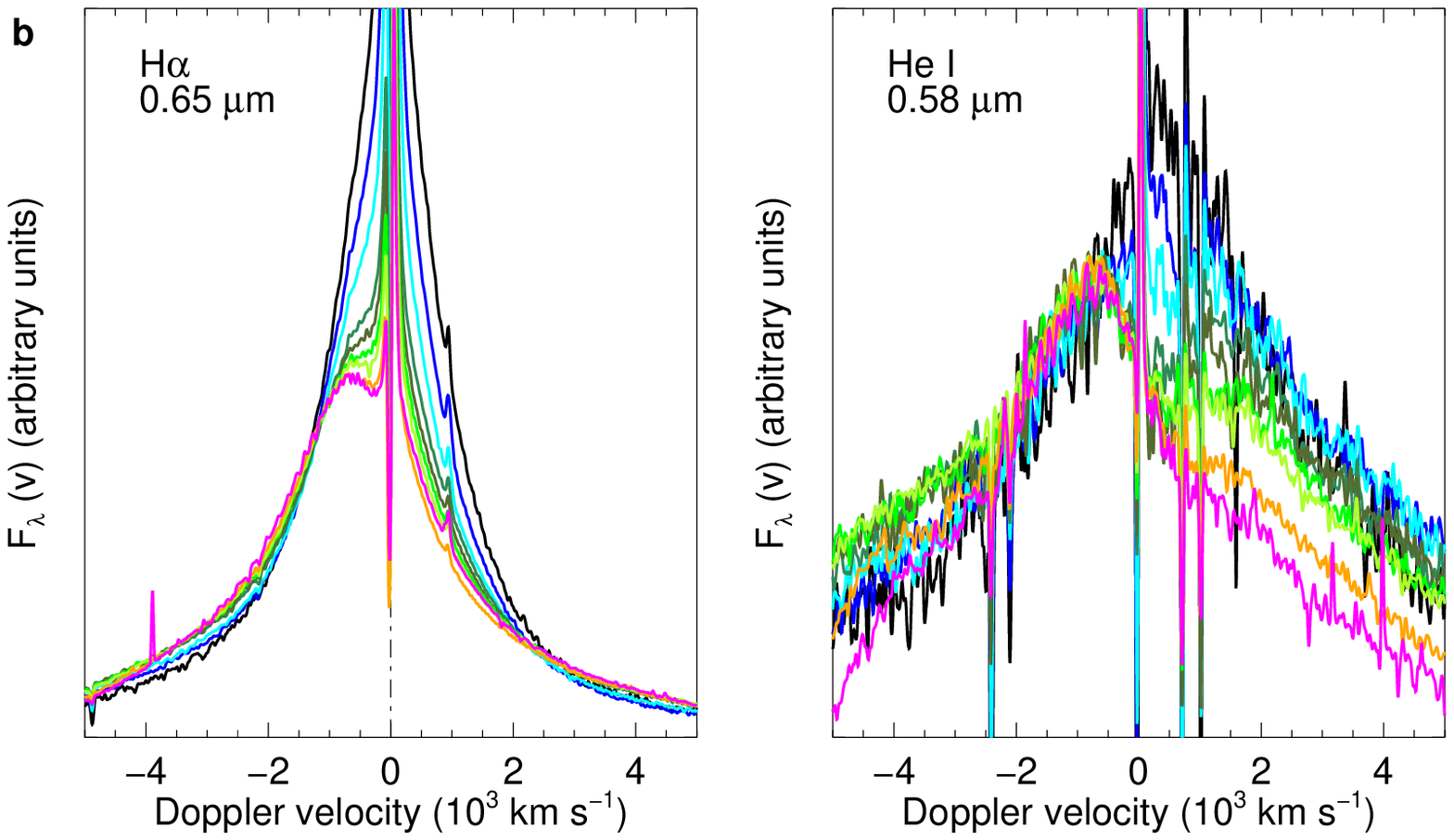}   
\end{center}
\vspace{-1.3 cm}
{\noindent\bf Extended Data Figure 3 $\mid$  
Line profiles.}    
{\bf a,} 
Comparison of the observed line profile 
(left panel) to the line profile of the Lorentzian line fits (right panel), 
illustrated for H$\beta$ $\lambda$4861.35. 
{\bf b,} Left panel: Line profile of the H$\alpha$ $\lambda$6562.79 line. 
The progressive broadening of the line causes both the blue and 
red wings to cross at different epochs.
Right panel: The line profile of the \ion{He}{i} $\lambda$5875.621 
line exhibits a similar effect.
The lines increasingly deviate from a Lorentzian profile. 
%
%
\newpage
\begin{center}
\leavevmode
\\
 \includegraphics[scale=0.88]{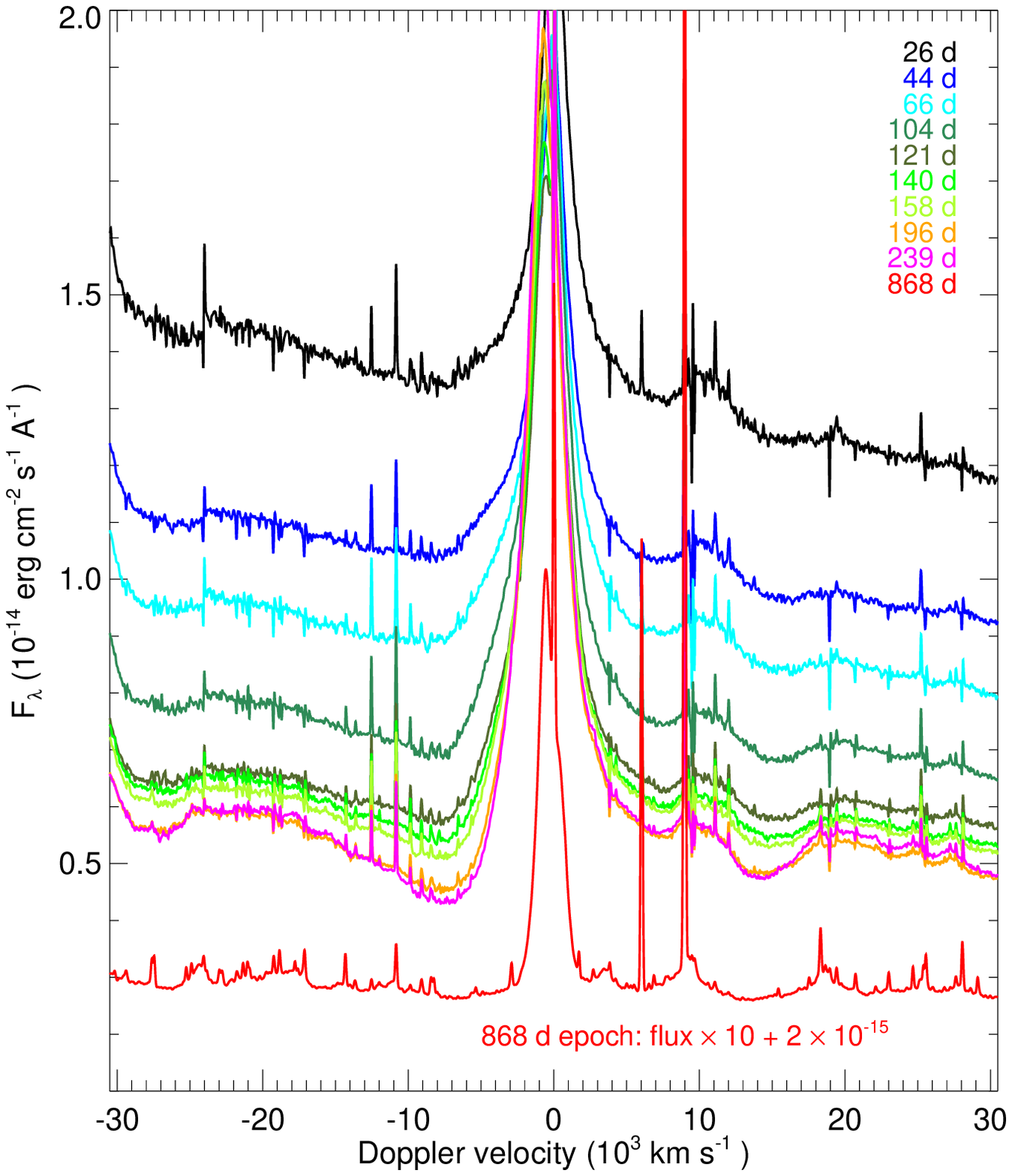}
\end{center} 
\vspace{-1.6 cm}
{\noindent\bf Extended Data Figure 4 $\mid$  
Development of the broad P~Cygni profile of H$\beta$.} 
Within the early epochs ($<$ 239 days) the hydrogen emission 
line H$\beta$ $\lambda$4861.35 develops a strong P~Cygni 
profile. The minimum of the P~Cygni profile is at $\sim$ 7,500 
km s$^{-1}$. The largest velocities associated with the P~Cygni 
profile are at $\approx$ 20,000 km s$^{-1}$.  
The late epoch (868 days) has been 
scaled by a factor 10 and offset for better comparison to the 
early epochs. The H$\beta$ line no longer exhibits 
features of high velocities.  
The wings of the intermediate velocity component extend to
$\sim$ 2,000--3,000  km s$^{-1}$. 
%
%
%
\newpage
\begin{center}
\leavevmode
\includegraphics[scale=0.75]{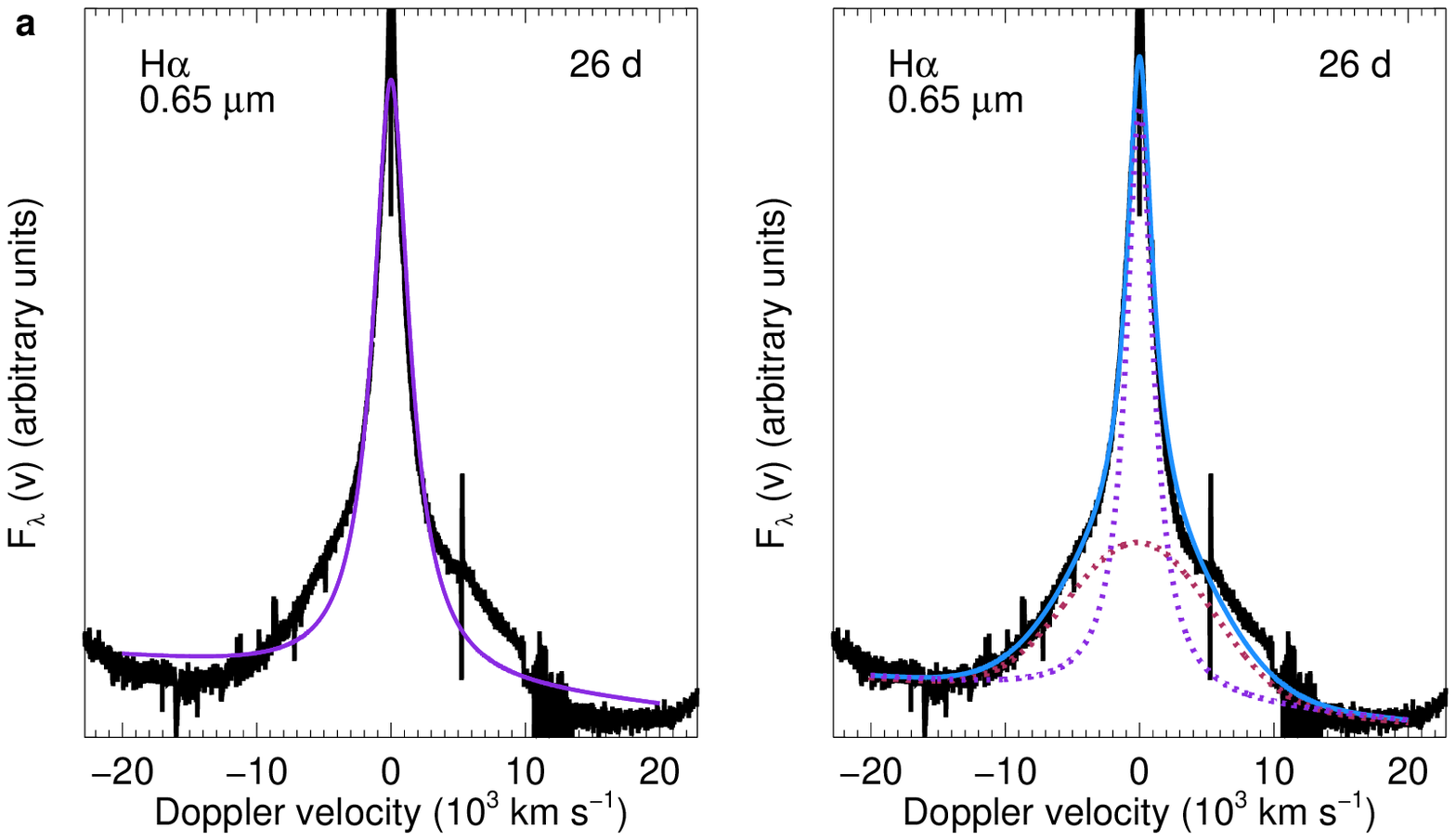}   
\includegraphics[scale=0.75]{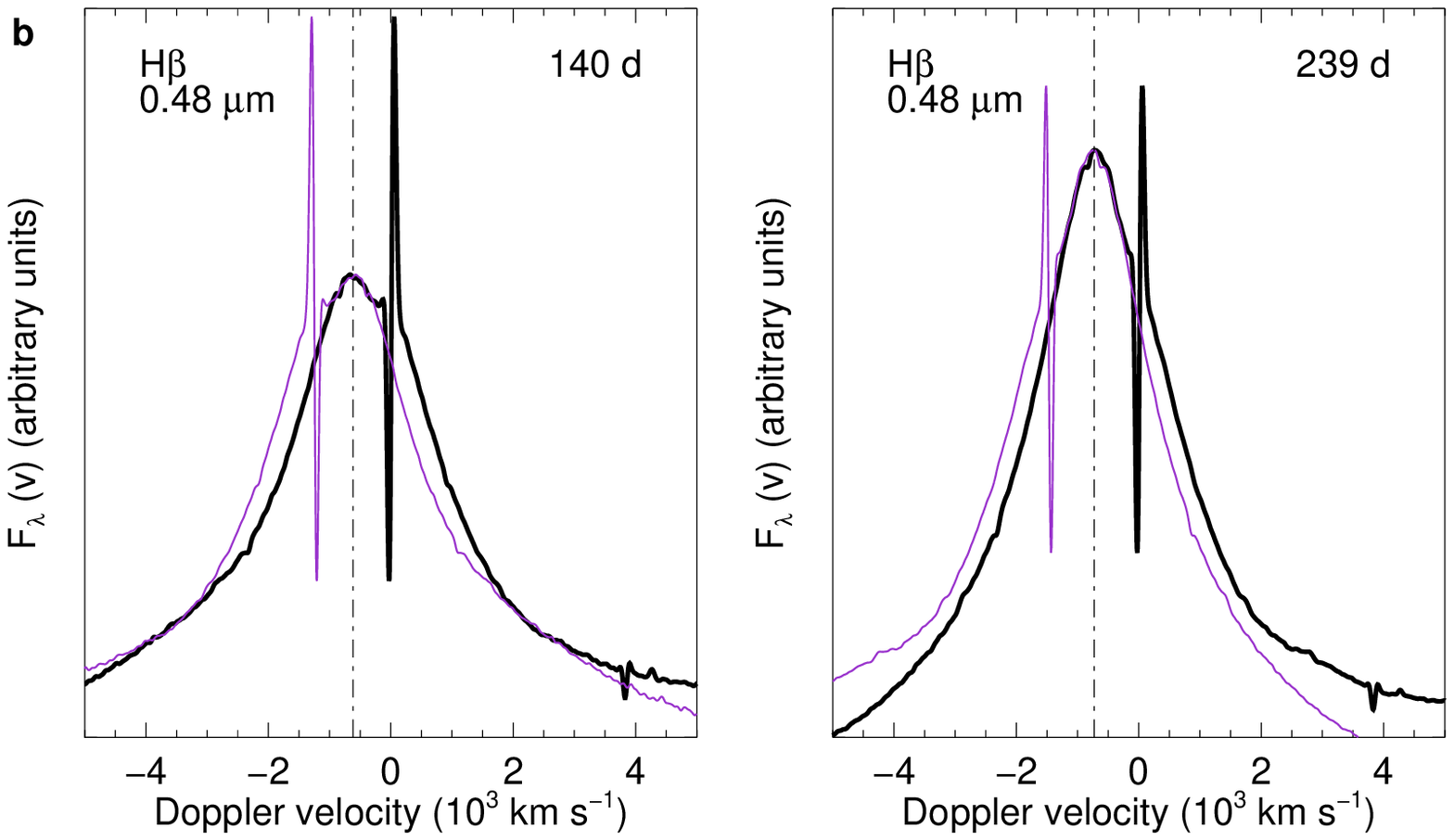}   
\end{center}
\vspace{-1.6 cm}
{\noindent\bf Extended Data Figure 5 $\mid$  
Velocity components and asymmetry 
of the intermediate emission lines.}    
{\bf a,}
Left panel: The H$\alpha$ $\lambda$6562.79 
line cannot be fit with a single Lorentzian 
(purple solid curve). Right panel: The broad 
(pink dotted curve) and the intermediate velocity 
component (purple dotted curve) and the 
combination of the two (blue solid curve).
{\bf b,} 
The H$\beta$ $\lambda$4861.35 line is 
asymmetric with respect to its peak velocities 
($\sim -$458 km s$^{-1}$ at 140 days and 
$\sim -$768 km s$^{-1}$ at 239 days). The 
mirrored emission lines are shown as purple 
thin curves.  The mirror axis is shown as black 
dashed-dotted curve.   Similar effects are seen 
for other emission lines.
%
%
%
\newpage
\begin{center}
\leavevmode
\includegraphics[scale=0.98]{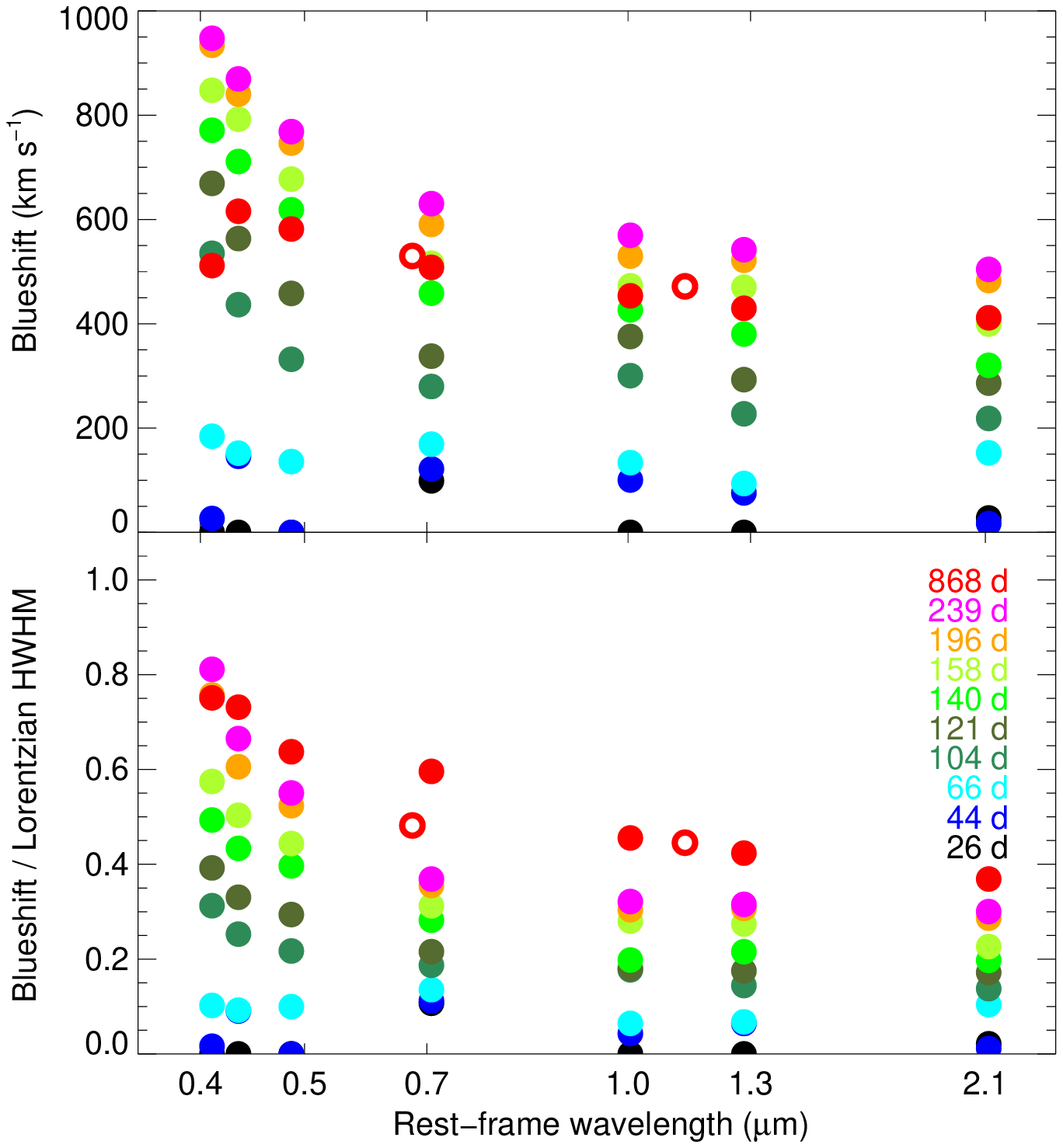}   
\end{center}  
\vspace{-2.0 cm}
{\noindent\bf Extended Data Figure 6 $\mid$  
Evolution of the blueshift velocity of hydrogen and metal lines.} 
The blueshift of the hydrogen lines is wavelength 
dependent and increases with time for the early epochs.
At any epoch the blueshift is smaller for lines at longer wavelengths.  
The filled symbols correspond to the blueshifts of the hydrogen 
emission lines and the open circles correspond to the oxygen 
lines.
The blueshift to HWHM ratio for the early 
epochs resembles the extinction curves (Figure 2).  
%
%
%
\newpage
\begin{center}
\leavevmode
 \includegraphics[scale=0.98]{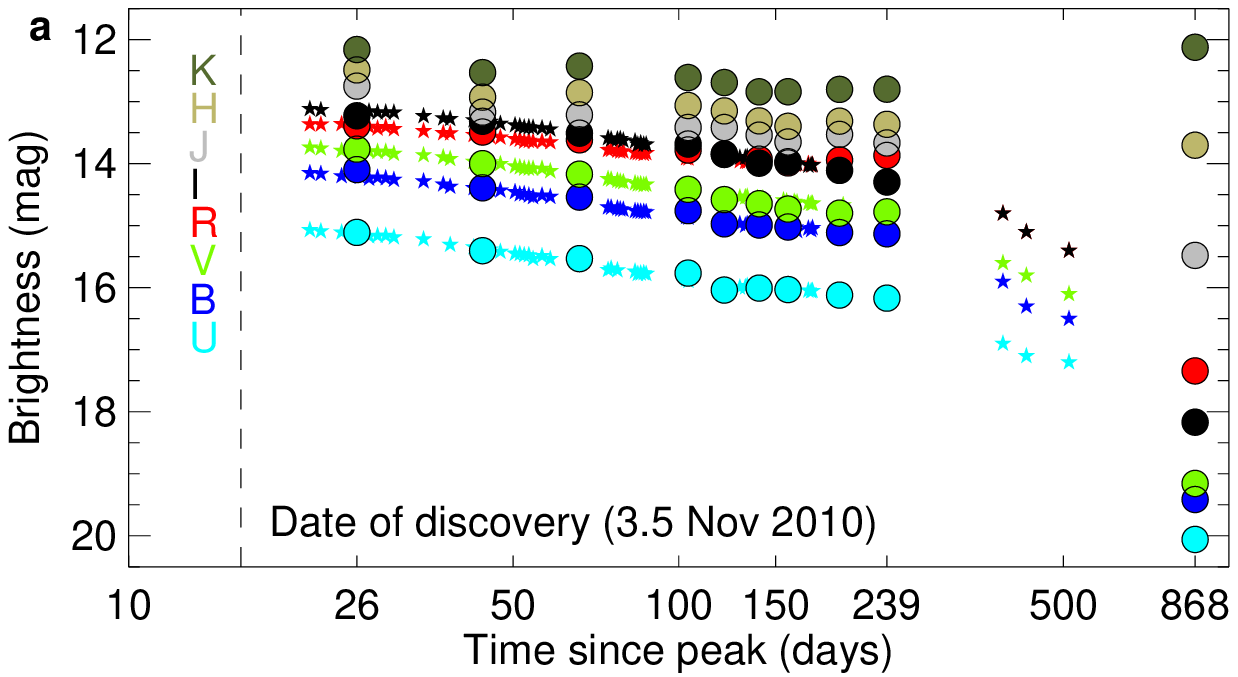}
  \includegraphics[scale=0.98]{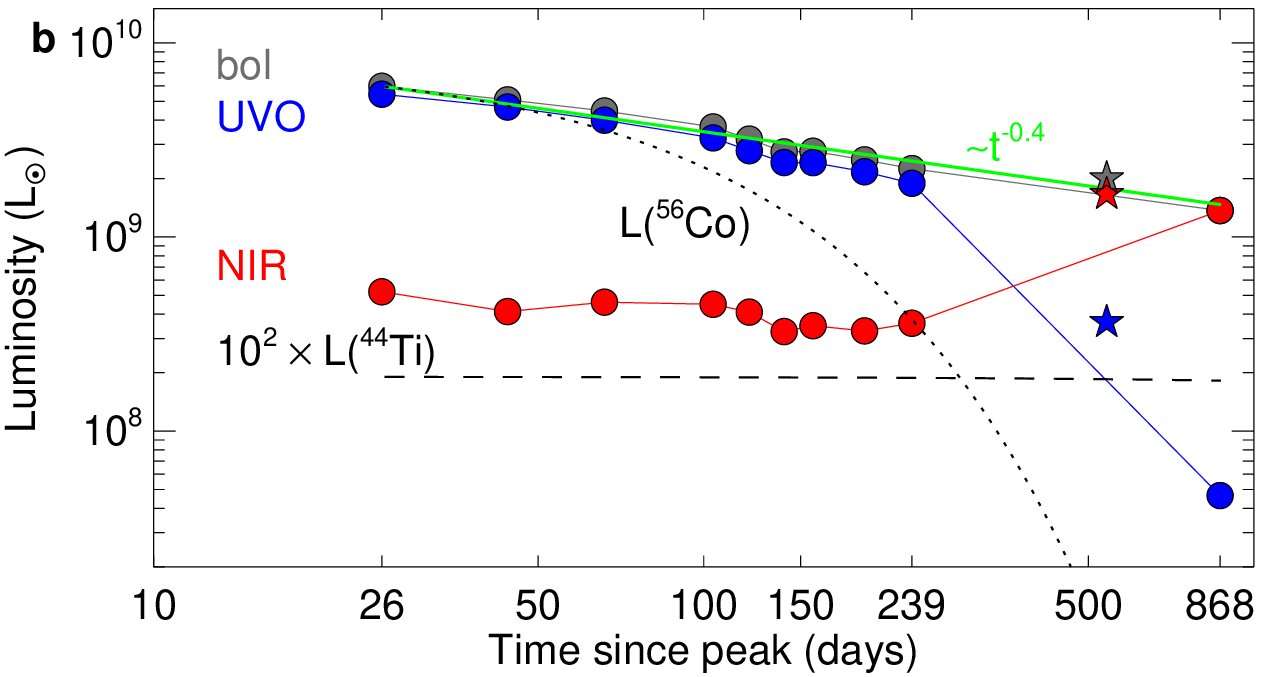}
\end{center} 
\vspace{-1.0 cm}
{\noindent\bf Extended Data Figure 7 $\mid$  
Lightcurves.} 
{\bf a,} Synthetic UVBRI and JHK
lightcurves (filled circles) compared to the UBVRI optical 
photometry of  \citemeth[Ref.][]{2013ApJ...776....5M} (small stars). 
{\bf b,} Energy output.
The temporal evolution of the UVO and NIR 
luminosities (blue and red symbols, respectively) and 
the total bolometric (UVO + NIR) luminosity (black diamonds). 
The green curve is a $t^{-0.4}$ power-law 
approximation to the UVO emission at early 
times. 
We have included data points 
from the literature (filled stars) at 553 days 
\citemeth{2013ApJ...776....5M}.  The
maximum possible contributions to the heating of the ejecta 
from the radioactively decaying  $^{56}$Co and the isotope $^{44}$Ti 
are shown as a dotted curve and a dashed line, respectively.
%
%
%
\newpage
\begin{center}
\leavevmode
\includegraphics[scale=1.0]{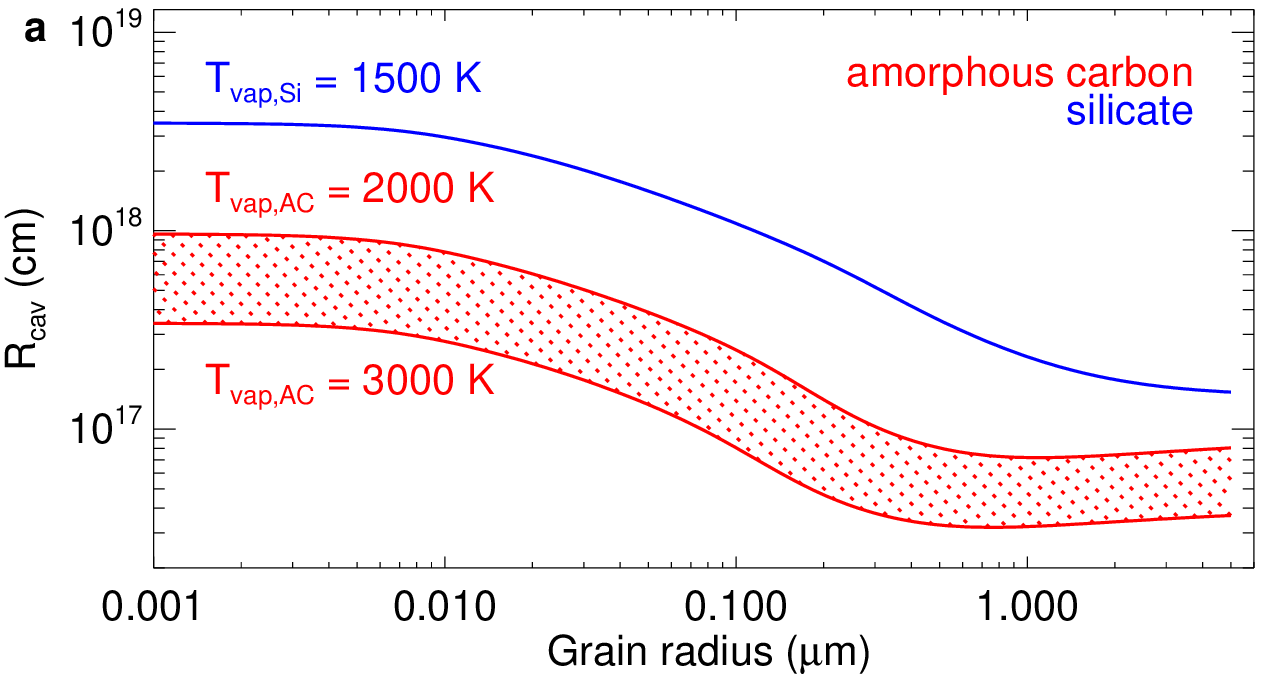}   
 \includegraphics[scale=1.0]{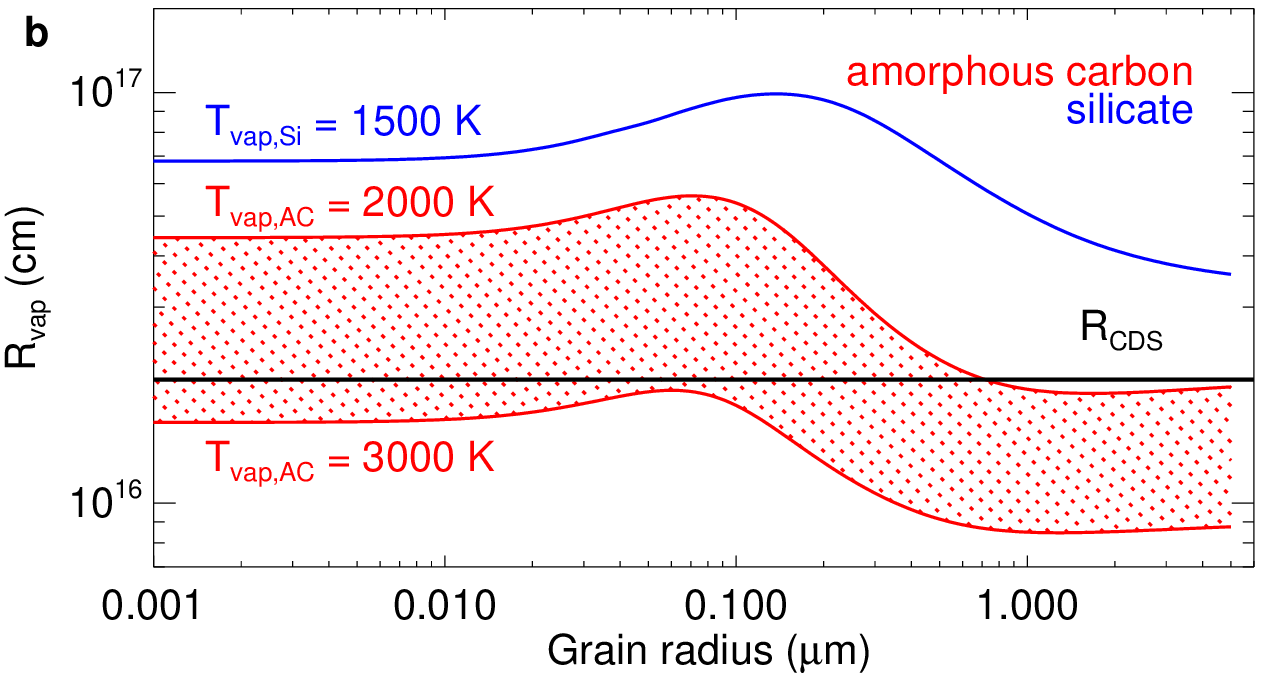}   
 \includegraphics[scale=1.0]{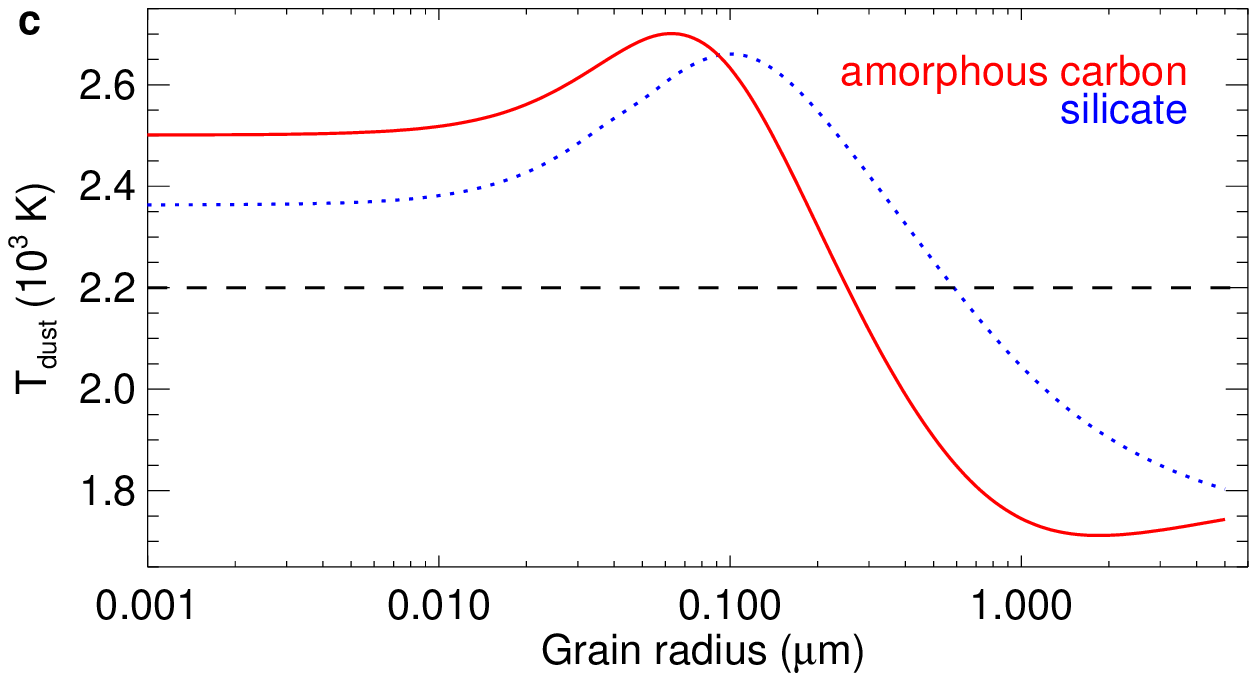}   
 \end{center}  
 
{\noindent\bf Extended Data Figure 8 $\mid$  
Dust vaporization radii and temperatures as a function of grain radius.} 
{\bf a,} Radii, $R_{\mathrm{cav}}$, from an initial burst of radiation.  
{\bf b,} Radii, $R_{\mathrm{vap}}$, from the observed supernova luminosity at 26 days.   $R_{\mathrm{cav}}$ and $R_{\mathrm{vap}}$ depend on the vaporization temperatures
$T_{\mathrm{vap,AC}}$ and $T_{\mathrm{vap,Si}}$. 
The black line indicates the location, $R_{\mathrm{CDS}}$, of the CDS.  
{\bf c,} 
The dust temperatures at $R_{\mathrm{CDS}}$, for grains
heated by the supernova light and cooled through the NIR emission. 
The dashed line indicates $T_{\mathrm{hot}}$ 
derived from the spectral fits (26 days).
Amorphous carbon grains (solid curve) have 
temperatures $\lesssim T_{\mathrm{vap,AC}}$.
Silicate grains (dotted curve) 
would be hotter than $T_{\mathrm{vap,Si}}$ and therefore cannot exist.
%
%
%
\newpage
\vfill
\begin{center}
\leavevmode
\includegraphics[scale=1.0]{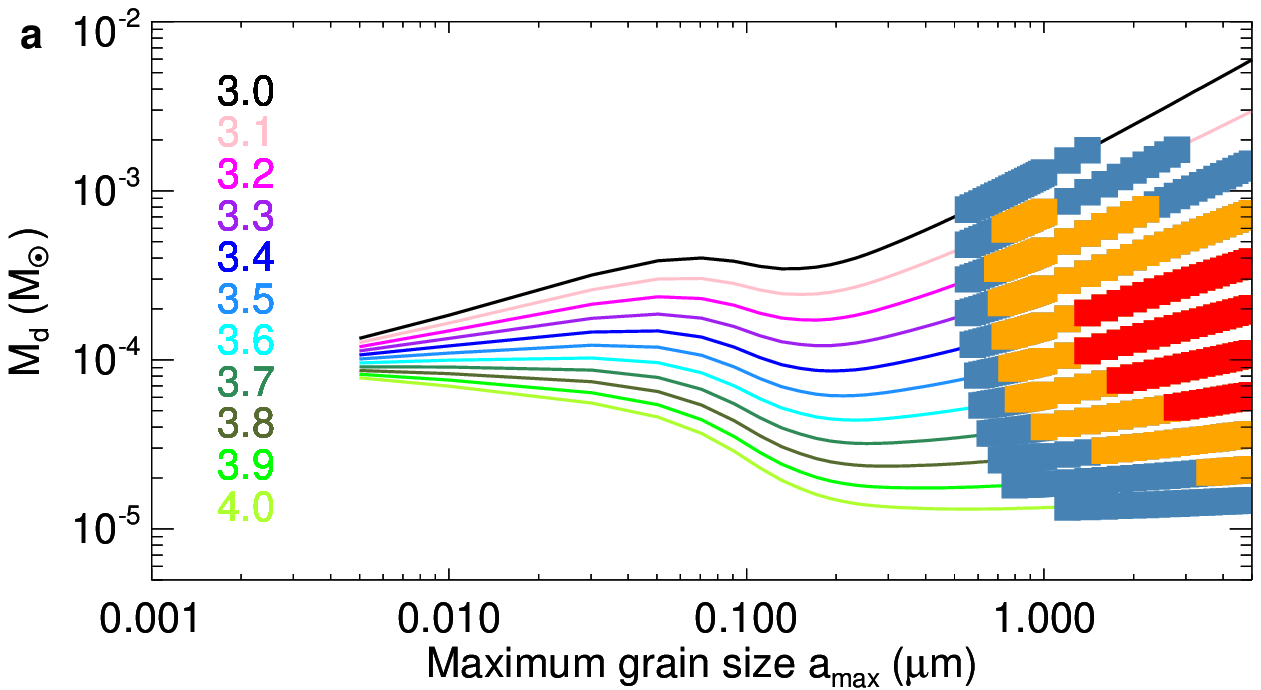} 
\includegraphics[scale=1.0]{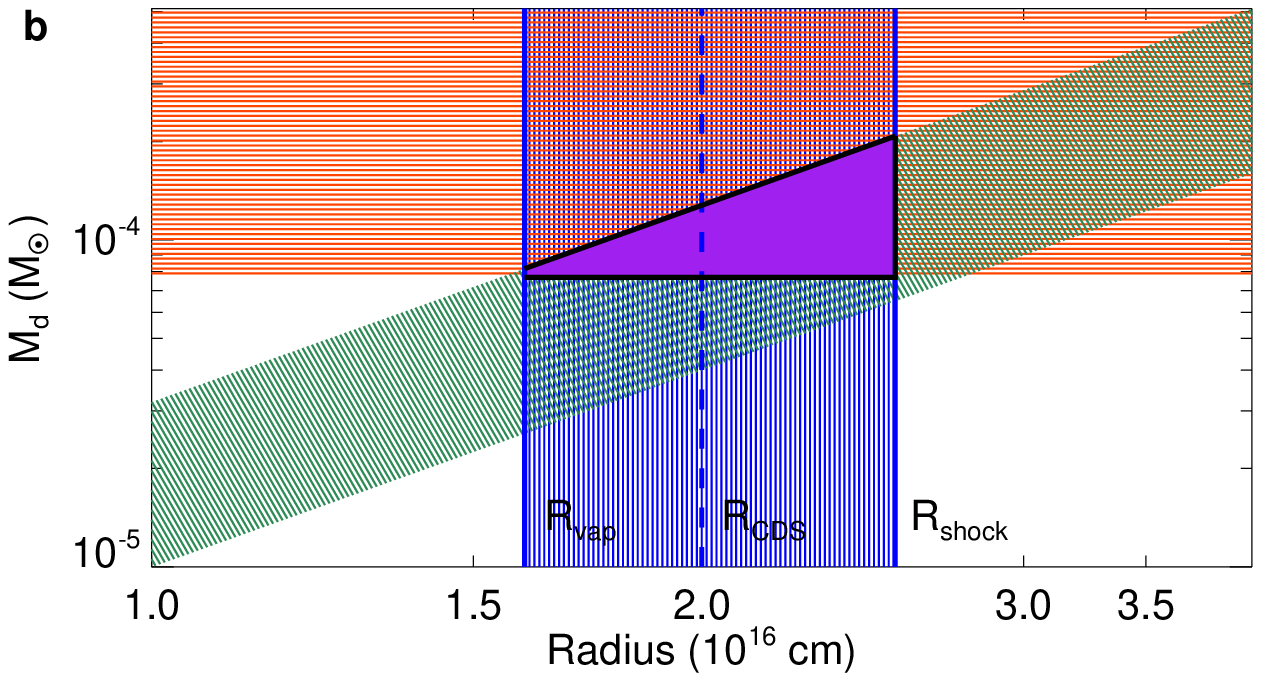}
\end{center} 
\vspace{-1.0 cm}
{\noindent\bf Extended Data Figure 9 $\mid$  
Dust mass at 239 days past peak.} 
{\bf a,} Sensitivity of the dust mass to the parameters 
$a_{\mathrm{max}}$ (coloured curves) and $\alpha$ of the grain 
size distribution function. The filled coloured 
squares represent the dust masses for the parameters of the 
grain size distribution function of the 1$\sigma$ (red), 
2$\sigma$ (orange) and 3$\sigma$ (blue) confidence interval
(Figures 2c and 3).
{\bf b,} 
The extinction dust mass and its 
standard deviation (green-shaded band), the 
dust mass from the NIR emission (red-shaded band)
and the radius range $R_{\mathrm{vap}}$ $\lesssim$ 
$R_{\mathrm{CDS}}$ $\lesssim$ $R_{\mathrm{shock}}$ 
(blue lines and shaded area). 
The overlapping region (purple framed 
area) of the three bands constrains 
$R_{\mathrm{CDS}}$ and the dust mass. 
%
%
%
%
%
\newpage
{\noindent\bf Extended Data Table 1 $\mid$  
 Log of the VLT/X-shooter observations of SN 2010jl.} 

\begin{table}     
\label{TAB:DCCSN}
\centering
\footnotesize
\begin{tabular}{lcccccc}
\hline
   Date (UT)       		
   &Airmass  	
   &Seeing ($''$) 
   & \multicolumn{3}{c}{Exposure times (s)}
   & Days past peak on \\

   & 
   & 
   &UVB
   &VIS
   &NIR 
   & 2010 Oct 18.6 UT\\ 
        
\hline

   2010 Nov 13.4		& 1.53	& 0.91	& 2$\times$100	& 2$\times$100	& 2$\times$100	 &26\\
   2010 Dec  1.4		& 1.25	& 1.74	& 2$\times$250	& 2$\times$250	& 8$\times$100	 &44\\
   2010 Dec 23.3		& 1.22	& 1.52	& 2$\times$250	& 2$\times$250	& 8$\times$100	 &66\\
   2011 Jan 30.3		& 1.21	& 1.05	& 2$\times$250	& 2$\times$250	& 8$\times$100	 &104\\
   2011 Feb 16.1		& 1.34	& 0.87	& 2$\times$250	& 2$\times$250	& 8$\times$100	 &121\\
   2011 Mar 7.2			& 1.24	& 0.77	& 2$\times$250	& 2$\times$250	& 8$\times$100	 &140\\
   2011 Mar 25.1		& 1.21	& 0.98	& 2$\times$400	& 2$\times$450	& 10$\times$100	 &158\\
   2011 May 3.0			& 1.21	& 0.74	& 2$\times$400	& 2$\times$450	& 10$\times$100	 &196\\
   2011 Jun 15.0		& 1.81	& 0.90	& 4$\times$550	& 4$\times$600	& 32$\times$100	 &239\\
   2013 Mar 4.0			& 1.28	& 0.86	& 8$\times$698	& 8$\times$605	& 56$\times$100	 &868\\

\hline\\
\end{tabular}\\
\end{table}

%
\newpage
\bibliographystylemeth{naturemag}
\bibliographymeth{sn2010jl}
%
\end{methods}
\newpage

\begin{supp}

\section{Origin of the lines and line attenuation} 

The different velocity components can be associated with 
different regions from which the emission arises.
The intermediate velocity component ($\sim$ 400--4,000 km s
$^{-1}$) in Type IIn supernovae is usually attributed to emission from a 
postshock region behind the forward shock, which penetrates 
the dense circumstellar medium  
(13, 18).        
The conspicuous narrow P~Cygni lines arise from interaction 
with a slow unshocked circumstellar stellar wind 
\citesup{2011ApJ...732...63S}.
The presence of rather strong intermediate and narrow width 
hydrogen lines suggests that SN 2010jl has a hydrogen rich 
circumstellar medium, but has not shed all the hydrogen during 
its progenitor phase, which can be deduced from the broad 
component of H$\alpha$ (Extended Data Figures~3b~and~5a). 

Our high-resolution data set clearly shows that the intermediate-
width emission lines (H, He) feature a progressive attenuation of 
their red wings. The attenuation effect is strongest for lines at 
the shortest wavelengths and gradually weakens towards lines 
at longer wavelengths (Figure~1). 
Furthermore, the red wing attenuation is accompanied by a 
corresponding wavelength dependent blueshift of the centroids 
of the lines (Extended Data Figure~6). These effects together 
with the temporally increasing attenuation and blueshifts point to 
ongoing dust formation in a shell like geometry, i.e., in a cool 
dense shell (CDS) behind the forward shock at the early epochs. 
The persistent blueshift and strong attenuation of the emission 
lines at the late epoch hint at dust formation in the ejecta 
(as the blueshift is detected in the ejecta metal lines).

Dust formation in the ejecta can be ruled out for the early 
epochs (26--239 days) because the ejecta are still too hot  to 
allow for dust condensation. Furthermore, emission lines of 
nucleosynthesized elements residing in the ejecta, which would 
show effects of dust formation in their profiles, are not detected. 
Thus, the majority of the ejecta material out of which dust can 
condense is still behind the optically thick and hot 
(T$_{\mathrm{SN}}$ $\approx$ 7,300 K) photosphere of the 
supernova (SN). 

It is unlikely that the observed progressive attenuation and 
blueshift of the lines is due to occultation. The opaque ejecta 
could in principle block the light from the far side of the ejecta, 
leading to an attenuation of red light (emitted by receding 
material). However, the area of the photosphere 
($R_{\mathrm{SN}}$ $\sim$ 2.5--3 $\times$ 10$^{15}$ cm)
is just about 2 percent of that of the CDS 
($R_{\mathrm{CDS}}$ $\sim$ 2 $\times$ 10$^{16}$ cm, see
Methods and Supplementary Information section 2) from which we assume that the 
intermediate emission lines originate, making this effect 
negligible. Additionally, a pure occulting effect would lead to a 
decreasing attenuation of the emission lines with time because 
of the receding photosphere, which is the opposite of what we 
observe. 

Alternatively, the photosphere is in front of the forward shock 
and the emission line velocities do not represent physical motion 
of the shock but are rather caused by Thompson electron 
scattering  
\citesup[13, 40,][]{2001MNRAS.326.1448C}.
As the photosphere recedes it will be located behind the forward 
shock and the CDS. At this point, the intermediate-width 
emission lines become less dominated by the Thompson 
electron scattering effect and their velocity widths correspond to 
the physical motion of either the forward shock or the CDS. 
The suggestion that the blueshifts and attenuations are obtained 
through electron scattering and occultation (14, 40)
is however 
difficult to maintain. First, a significant effect due to electron 
scattering and occultation would require the lines to be formed 
at the photosphere \citesup{2009MNRAS.394...21D} which is ruled 
out by polarimetric measurements \citesup{2011A&A...527L...6P} 
showing significant depolarization of the intermediate width 
lines. Second, the wavelength dependent effect which we 
observe (Figure~1, Extended Data Figure~6), 
in particular the fact that the $n=5$ transitions H$\gamma$ and 
P$\beta$ show different blueshifts, is difficult to attribute to
electron scattering. 
Moreover, in this picture, the emission lines are expected to be 
symmetric, resembling Lorentzians. Although the hydrogen lines 
used in our studies are well approximated by Lorentzian profiles, 
they are not symmetric as illustrated in Extended Data 
Figure~5b. Some hydrogen emission lines have been excluded 
from our studies due to significant asymmetries and deviations 
from Lorentzian profiles (Methods). The emission lines at the 
late epoch (868 days) can no longer be approximated by 
Lorentzian profiles. 
Additionally, we detect significant broad components
\citesup[Extended Data Figures~3,~4 and~5a, 
see also Ref.][]{2011A&A...527L...6P} which would not
be visible if the shell is optically thick. These broad lines suggest 
large velocities which would quickly overtake the CDS
(Extended Data Figure~4). Furthermore, we infer a black body 
radius, $R_{\mathrm{SN}}$, which is temporally receding 
(Methods). We attribute this to the 
receding photosphere rather than an optically thick shell at the 
shock interaction region. Such an optically thick shell would be 
expected to either remain at constant radius or increase with 
time due to the expanding forward shock.   

Finally, the suggestion (12) 
that there is a 
large difference in the blueshifts between the blue (H$\gamma$) 
and red (Br$\gamma$) is not supported by our observations. We 
attribute this conclusion to the low signal to noise ratio and the 
low resolution of the spectra upon which that conclusion was 
reached.
%
%
\section{Constraining the locations and sources of dust formation}

To determine the location of the CDS we require 
$R_{\mathrm{vap}}$ $\lesssim$ $R_{\mathrm{CDS}}$ $\lesssim$ 
$R_{\mathrm{shock}}$ and consistency between the extinction 
and emission dust masses. At 26 days, the blast wave reaches 
a radius $R_{\mathrm{shock}}$ $\gtrapprox$ 2 $\times$ 10$^{16}$ cm
for an assumed fast initial blast wave velocity of 
3.5 $\times$ 10$^{4}$~km~s$^{-1}$. 
As the SN blast wave decelerates to a velocity of about 
$\sim 3,000$
km s$^{-1}$, a CDS forms at a location 
$R_{\mathrm{CDS}}$ $\lesssim$ $R_{\mathrm{shock}}$ (both of 
these radii increase with time). 
Any dust must be located beyond the vaporization radius, 
$R_{\mathrm{vap}}$, of the nearly constant observed SN 
radiation. From Extended Data Figure 8 we find that (at 26 days)
$R_{\mathrm{vap}}$ $\lesssim$ 2 $\times$ 10$^{16}$ cm for 
carbonaceous dust with grain radii $\gtrsim$ 0.25 $\mu$m and 
temperature $T_{\mathrm{dust}}$ $\lessapprox$ $T_{\mathrm{hot}}$ 
while grains $\lesssim$ 0.25 
$\mu$m have higher temperatures of 2,200--2,700~K.
The calculated 
temperatures of the large dust grains at that location is 
consistent with those inferred from the modified black-body fits 
to the NIR emission.

Extended Data Figure~9b 
shows that the derived radius-dependent extinction dust mass, 
the dust mass from the NIR emission, and the characteristic radii of 
the system, $R_{\mathrm{vap}}$ and $R_{\mathrm{shock}}$,  
lead to a joint constraint on the location of the dust formation site 
of $R_{\mathrm{CDS}} \approx 2.0 \times 10^{16}$ cm.
An alternative location for early dust formation between 26--239 
days, such as the SN ejecta, does not meet the above 
requirements and is therefore disfavoured.

It has been pointed out (14)	
that there is  very little colour 
evolution, implying little extinction along the line 
of sight to the photosphere. This suggests that the dust formed 
outside the line of sight, e.g., in a non-spherical geometry or in 
an inhomogeneous CDS. The light from the region emitting the 
intermediate velocity component emission lines traverses the 
dust forming CDS, which does give rise to the extinction of the 
lines. Alternatively, a clumpy medium can give rise to a flat 
extinction curve \citesup{1997AJ....114..107B,
1999AJ....117.2226S}. In our case, however, the constancy of 
the extinction curve with time rules out that clumpiness is the 
cause of the flatness unless the increase in extinction is 
balanced by a corresponding change in the covering fraction.

At the late epoch (868 days) the 
inferred carbon dust mass from the NIR emission is more than 
10 times larger than the dust mass at 239 days (Figure~4).
The temperature of the dust has dropped 
to about 1,100~K and a larger black-body radius of 
$\approx 5.7 \pm 0.2 \times 10^{16}$ cm 
is inferred from fits to the NIR 
emission (Extended Data Figure~2). The CDS, which has 
a typical velocity of $\lesssim 3,000$ km s$^{-1}$, as inferred 
from the intermediate velocity component emission lines 
(Methods, Supplementary Information sections 1 and 4), 
will not have reached such a large radius. Moreover, 
the bulk ejecta material traveling with an average expansion velocity 
$v_{\mathrm{ej}} \approx$ 7,500 km s$^{-1}$, as measured 
during the first 236 days (Extended Data Figure~4), will have 
caught up with the CDS already.  This disfavours continued dust 
formation at an accelerated rate in the CDS. Instead, an 
additional dust source must emerge between 239 days and 868 
days.

We favour dust formation in an interaction region of the ejecta 
bulk material with swept up CSM. The bulk material of the ejecta 
will have reached a large radius corresponding to the derived 
black-body radius at 868 days ($(5.7 \pm 0.2) \times 10^{16}$ cm). 
This interpretation is supported by the highly attenuated and 
blueshifted hydrogen and oxygen emission lines (Figure~1).
Such an effect can only occur when the dust 
formation region resides inside or at the line emitting region. 
Furthermore, the oxygen lines are likely ejecta lines and thus 
strongly suggest that ejecta material is involved in the dust 
formation at late times. The supply of large amounts of dust 
forming ejecta material can explain the sudden acceleration in 
the build-up of the dust mass (Figure~4).

At late times the UBVRI optical lightcurves significantly drop (12, 24).
The synthetic photometry of the late epoch (868 days) shows a 
continued drop in the optical bands while the  H and K band 
lightcurves increase. This is reflected in the temporal evolution 
of the UV-optical (UVO) and NIR energy output shown in 
Extended Data Figure~7b. The drastic UVO drop has been 
attributed to dust formation (12), 
presumably at the onset of the accelerated dust formation 
detected here (Figure 4). On the other hand, we find that the evolution 
of the total bolometric luminosity continues to follow a $t^{-0.4}$ 
power-law, such as that found for the UVO lightcurve before the 
drop (24).
While at early times, the NIR 
emission is caused by the newly formed dust in the CDS, the 
dramatic change in the evolution of the light curves after day 
$\sim 236$, suggests a different origin for the NIR emission at 
late epochs ($\gtrsim$ 240 days). This may be due to
newly formed dust including ejecta material. At this stage,
the UVO luminosity from the shock interaction must be very 
efficiently absorbed and reradiated by the forming dust, 
providing an important constraint on any dust formation or 
evolution mechanism.
 
We propose a two-stage dust-formation mechanism. In the first 
stage, only CSM material is responsible for the formation of dust 
and in the second stage CSM and ejecta bulk material are 
involved. This is consistent with the evolution of the total 
bolometric luminosity (Extended Data Figure 7) which follows a 
$t^{-0.4}$ power law behavior up to 868 days. However, this
implies that the SN did not yet enter the snow-plow 
phase (conservation of momentum; the radiated energy behind 
the shock is comparable to the energy of the explosion)
of its evolution on day 868, contrary to suggestions in the 
literature (24).	
Presumably, the UVO 
luminosity is powered by a central light source, possibly 
by the interaction at the low radius of 
$R_{\mathrm{SN}}$ $\sim$ 2.4--3.2 $\times$ 10$^{15}$ cm 
between an aspherical, very high density SN progenitor 
outflow and the blastwave and ejecta material, which leads to 
optically thick emission. As mentioned above, the fact that
the early dust formation in the CDS at 
$R_{\mathrm{CDS}}$ does not cause substantial extinction of 
the UVO light suggests that dust initially forms outside the line of 
sight. The drastic drop in the UVO lightcurve (Extended Data 
Figure~7) is then due to extinction of the 
UVO radiation directly in the line of sight by the efficiently 
forming dust with the aid of ejecta material at large radius 
($(5.7 \pm 0.2) \times 10^{16}$ cm). 
The central light source will heat the 
dust at large radius to the observed temperature of $\sim$ 1,100 
K and cause the simultaneous rise in the NIR luminosity. 
We want to stress that the embedded heating sources from the radioactive 
decay of $^{56}$Co and the isotope $^{44}$Ti cannot power the 
observed NIR luminosities in SN 2010jl (Extended Data Figure~7). 
The geometry may be reminiscent of that of $\eta$~Car, which 
shows several very dense circumstellar arcs out to a few times  
10$^{16}$ cm \citesup{2013ApJ...773L..16T}.
%

%
\section{Ruling out pre-existing dust}  
The spectra exhibit narrow absorption lines from the resolved 
Na I doublet arising from absorption by material at the redshift of 
the SN as well as from the interstellar medium in the MW. We 
measure a total Na ID equivalent width (EW) of 
$0.354\pm0.004$ \AA~at the redshift of the SN. This translates 
into a host galaxy $E(B-V) = 0.037\pm 0.007$ mag
\citesup{2012MNRAS.426.1465P}, i.e., $A_V = 0.11\pm0.02$ mag 
for $R_V=3.1$. For the MW foreground we measure EW 
$=0.166\pm0.004$ \AA~from the Na ID lines, which corresponds 
to $E(B-V) = 0.022\pm0.004$ mag, i.e., $A_V = 0.07\pm0.02$ 
mag. This is consistent with the value obtained from maps of 
dust IR emission (37) 
used to correct 
our spectra for foreground extinction ($E(B-V) = 0.027$ mag) 
and the most recent recalibration ($E(B-V) = 0.024$ mag) based 
on Sloan Digital Sky Survey stellar spectra 
\citesup{2011ApJ...737..103S}. The combined host+MW $E(B-V)$ 
is $0.06 \pm 0.01$ mag. The Serkowski law 
\citesup{1975ApJ...196..261S}, $P_{\mathrm{max}} = 9E(B-V)$ \%, 
implies $P < 0.6$ \%, consistent with the measured $P < 0.3$ \% 
for the SN intermediate line emission \citesup{2011A&A...527L...6P}. 
The Na ID and polarization measurements set strong 
constraints on the reddening of the SN and hence on the total 
dust column along the line of sight. 

The lack of foreground dust extinction must be reconciled with 
the very high metal column density inferred from X-ray 
absorption observations. SN 2010jl was observed with 
{\it Chandra} at two epochs, on 2010 Dec 7 UT and 
2011 Oct 17 UT \citesup{2012ApJ...750L...2C}. At the first epoch a 
very large equivalent hydrogen column density of 
$1\times 10^{24}$ cm$^{-2}$ was inferred from soft X-ray 
absorption. At the second epoch this had dropped to 
$3\times 10^{23}$ cm$^{-2}$.  Applying a standard MW 
conversion factor between $A_V$ and $N_{\mathrm{H}}$ 
\citesup{2011A&A...533A..16W} and accounting for the assumed 
metallicity  \citesup{2012ApJ...750L...2C} this would correspond to 
$A_V$ of 136 and 40 mag, respectively. Clearly SN 2010jl was 
not extinguished by this much dust, confirming that most of any 
pre-existing dust must have been vaporized by the SN out to 
radii larger than $R_{\mathrm{cav}}$ of about 10$^{17-18}$ cm
(Methods).

Furthermore, from Extended Data Figure~1 it is evident that 
while the {\em Spitzer}/IRAC 3.6 $\mu$m observations of SN 
2010jl (11) 
are well-reproduced by our 
modified black-body fit to the X-shooter data, there is significant 
excess emission in the 4.5 $\mu$m band. This has been 
interpreted as being due to a cold component of pre-existing 
dust emitting at 750~K in a torus around the SN (11).
However, the model upon which 
that claim was reached cannot account for the NIR excess in 
the range of 1--2.3 $\mu$m, evident in all our spectra 
(Extended Data Figures~1 and~2). As demonstrated in our 
analysis, a hot dust component of $\sim$ 1,800 K is required to 
account for the NIR excess and simultaneously for the 
3.6 $\mu$m emission. Additionally, as shown in 
Extended Data Figure~8, no dust with temperatures 
$\lesssim$ 1,500~K can exist at the location of the CDS. Thus, 
any such cool dust must be placed outside the dust free cavity 
at radii $\gtrsim$ $R_{\mathrm{cav}}$, which we disfavoured 
based on the extinction constraints presented above. 

Alternatively, the remaining excess 4.5 $\mu$m emission could 
be interpreted as due to 4.7 $\mu$m fundamental band 
emission of carbon monoxide (CO). Following the first 
detection of CO (fundamental band and first overtone at 
2.3 $\mu$m) in SN 1987A \citesup{1988Natur.334..327S,
1988MNRAS.231P..75C,1988slmc.proc...37D}, CO has been 
detected in numerous Type II SNe, including the detection of 
the first overtone in the Type IIn SN 1998S 
\citesup[see Ref.][for an overview]{2011IAUS..280..228C}. Excess 
4.5 $\mu$m band photometric excess has previously been 
attributed to the CO fundamental band in Type II SNe, e.g., in 
SN 2007it \citesup{2011ApJ...731...47A}. CO is an abundant 
molecule which forms at very high temperatures (up to $\sim$ 
5,000~K), and acts as strong coolant through its 
rotational-vibrational transitions. Thus, it is considered to significantly 
influence subsequent dust formation 
\citesup{2001MNRAS.325..726T,2003ApJ...598..785N,
2011IAUS..280..228C}. The possible detection of CO adds to 
the evidence that conditions are right for the early formation of 
dust in SN 2010jl. 

The hot NIR emission of the early epochs cannot be attributed 
to a thermal echo generated by pre-existing dust heated by a 
short burst of far UV radiation associated with the shock 
breakout. Such a burst  will have vaporized any pre-existing dust in the 
CSM out to about 10$^{17-18}$ cm
(Extended Data Figure~8),
much larger than the NIR black-body radius. 
The surviving dust will reradiate the absorbed energy from the 
burst, giving rise to a NIR echo with temperatures between 
1,500 to 2,000~K, depending on dust grain radii and composition. 
The locus of the emitting dust describes an ellipsoid of 
revolution with a thickness $\lesssim 1$~light day, with one 
focus centred on the SN and the other on the observer. At time 
$t$, the radiating dust lies on the intersection of the ellipsoid 
with the (assumed) sphere of surviving dust, creating an 
annulus with a projected area $A \approx 2\pi c^2 t \Delta t$. 
The total mass of radiating dust constitutes a fraction equal to 
$A/4\pi (ct)^2 = \Delta t/2t$ of the surviving circumstellar dust. 
On day 66 past peak (assuming this is about $t \approx 100$~d 
past explosion) the mass of hot dust observed in the SN 
spectrum is about 
$M_{\mathrm{hot}} \approx 5 \times 10^{-5}$~M$_{\odot}$. If 
attributed to an echo, the total mass of dust in the shell must be 
$2t/\Delta t \approx 200$ times larger. 
The visual optical depth of such shell is then 
$\tau_{V}$ = $\tau(\lambda = 0.55 \mu$m) 
$ \approx 200\times (M_{\mathrm{hot}}/4 \pi R_{\mathrm{cav}}^2) \kappa_{V}$. For
 $\kappa_{V} \approx 3\times 10^4$~cm$^2$~g$^{-1}$, the 
visual optical depth $\tau_{V}$ $\approx$ 0.5. This value is 
significantly larger than the observed upper limit of $\tau_{V}$ 
$ \lesssim 0.1$. Thus, the large mass of surviving dust needed 
to give rise to an echo will violate the observed upper limits on 
the amount of intervening extinction, ruling out a NIR echo (40)
as the source of the early hot NIR 
emission.
At early epochs, the dust emission therefore most likely 
originates from inside the vaporization zone, where the 
subsequently forming dust gives rise to both the extinction and 
the NIR emission. This supports the constraints on the dust 
formation location derived in Supplementary Information 
section~4. 

The NIR emission at late times ($>$ 300 days) has been 
attributed to a dust echo from pre-existing dust (40).
We find that at late epochs, the observed NIR emission cannot 
be attributed to an echo from pre-existing dust, which escaped 
vaporization by the shock breakout. In this picture, the break in 
the UVO luminosity is not caused by extinction. This would 
suggest that the SN shock has entered the snowplow phase of 
its evolution after day $\sim 300$, or propagated beyond the 
outer boundary of the CSM. The SN light curve can then 
be thought of as a top-hat function with an average luminosity 
of $\sim 3\times 10^9~\Lsun$ and a width of $\sim 250$ days. 
Furthermore, the dust could not be distributed in a shell, since 
we would then expect a NIR echo of the same intensity as the 
flux on days 553 (12) and 868 at earlier 
epochs as well, which is contrary to our observations. The dust 
must therefore be distributed in a torus with an inner radius of 
275~light days, to give rise to the observed flux on day 553.  
The cross-sectional radius of the torus may be much smaller 
than the width of the SN lightcurve ($\sim 250$~light 
days), in which case the echo on day 553 is caused by the 
encounter of the ring with the leading edge of the SN 
lightcurve, and that on day 868 by its trailing edge. The dust 
temperature is expected to be only about 600~K, since the UVO 
flux incident on the dust is smaller by a factor of $\sim 10^3$ 
than that incident on the dust heated by the shock breakout 
($L \approx 10^{11}~\Lsun$, Methods)
to $\sim$ 2,500~K at a distance of 40~light days 
(Extended Data Figure~8). An echo origin for the 
NIR emission observed on days 553 and 868 can therefore be 
definitely ruled out.
Finally, the NIR luminosity on day 868 is about one third of the 
average UVO luminosity during the $\sim$30--300 day interval 
(Extended Data Figure~7), an excessively high fraction of 
the UVO luminosity to be intercepted by a dust torus. 

%
\section{Properties of the early dust formation site} 
%
The rapid formation of large dust grains requires a gas density 
of order $\sim$ 10$^{9-11}$ cm$^{-3}$ and temperatures below 
$\sim 3,000$~K. The CDS is therefore considered as the likely 
dust formation site \citesup[see Ref.][for a simulation of a forming 
CDS]{2010MNRAS.407.2305V}.
We can estimate the density of the dust-forming region from the 
X-ray observations that were conducted at 50 and 364 days 
past peak \citep{2012ApJ...750L...2C,2013ApJ...763...42O}. 
The observations suggest a gas temperature of about 8--10 
keV ($\sim$ 10$^8$~K). For a strong shock, the postshock gas 
temperature $T_{\mathrm{gas}}(K) = (3/16) \, k^{-1}  \, \mu \, v_{\mathrm{s}}^{2}$,
where $v_{\mathrm{s}}^{2}$ is the velocity of the forward shock 
propagating through the dense shell, $k$ is the Boltzmann 
constant and $\mu$ is the molecular weight of a fully ionized 
gas \citesup[see Ref.][]{2011piim.book.....D}. For a 
$T_{\mathrm{gas}}$ corresponding to the (X-ray) gas 
temperature, we infer a shock velocity of $\sim 3,000$~km~s$^{-1}$, 
as also found in Ref (24).	
The first evidence for dust formation was present in the spectra 
obtained 44 days past peak. We assume that the shock had 
collided with the dense shell shortly before our first observation, 
which requires that the gas had to cool within less than $\sim$ 
18 days from $\sim$ 10$^8$ to 3,000~K. 

The cooling time of a hot plasma is given by
\begin{equation}
 \tau_{\mathrm{cool}} = { k T_{\mathrm{gas}} \over n_e\, \Lambda(T_{\mathrm{gas}})},
\end{equation} 
where $n_e$ is the electron density of the postshock gas and 
$\Lambda(T_{\mathrm{gas}})$ is the cooling function. For X-ray 
emitting plasmas, 
$\Lambda(T_{\mathrm{gas}}) \approx 2\times 10^{-23}$~erg~s$^{-1}$~cm$^3$ 
at temperatures of $\approx 3~\times~10^{7}$~K, the `bottle 
neck' temperature where cooling is least efficient 
\citepsup{2009A&A...498..915D,2011piim.book.....D}.
We estimate the density of the postshock gas from the 
observed change in the hydrogen column density of the 
preshock shell between the two X-ray observations, 
$\Delta N_{\mathrm{H}}$ = 7 $\times$ 10$^{23}$ cm$^{-2}$, 
which is attributed to the passage of the forward shock through 
the shell. The average density of the preshock shell is then given by
$n_{\mathrm{H}} = \Delta N_{\mathrm{H}} / (v_{\mathrm{s}} \, \Delta t)$ $\approx$ 10$^{8}$~cm$^{-3}$, taking 
$\Delta t = 314$~d as the time interval between the two X-ray 
observations. For a strong shock the corresponding 
postshock density is 4 $n_{\mathrm{H}}$, which results in a 
cooling time of $\tau_{\mathrm{cool}}$ $\approx$ 6 days 
according to equation (1).
Once the gas has cooled below $\sim 3\times10^7$~K, rapid 
cooling via atomic line transitions will further compress and cool 
the gas \citesup{2011piim.book.....D}. Using the approximation of 
an `isothermal shock' we find that the density reached in the 
CDS will be $\gtrapprox$ 10$^{10}$ cm$^{-3}$ for the required 
temperature of the CDS. Similarly, if we assume that all the 
swept up material will be compressed into the CDS, we derive a 
density of the CDS from the X-ray observations as
$n_{\mathrm{H,CDS}}$ = $\Delta N_{\mathrm{H}}$ / $\Delta R_{\mathrm{CDS}}$ 
$\sim$ 5 $\times$ 10$^{9-10}$ cm$^{-3}$ for an adopted 
thickness of the CDS
$\Delta R_{\mathrm{CDS}}$ = 1--10 AU \citep{2010MNRAS.407.2305V}. 
We can approximate the minimum mass of the CDS from the 
swept up material between the two X-ray observations as 
$M \approx 4\pi m_{\mathrm{H}} \,R_{\mathrm{CDS}}^{2} \, n_{\mathrm{H}} \, v_{\mathrm{s}} \, \Delta t$ = 
$4\pi m_{\mathrm{H}} \, R_{\mathrm{CDS}}^{2} \, \Delta N_{\mathrm{H}}$ $ \approx$  3 $\Msun$, 
where  $R_{\mathrm{CDS}}$ = 2 $\times$ 10$^{16}$ cm is the 
assumed location of the CDS (Extended Data Figure~9b). The 
corresponding amount of metals is $\approx$ 0.02 $\Msun$. 
Furthermore, we estimate an accretion timescale 
\citesup{2011ApJ...727...63D} for dust grains with grain radii 
$a$ $\lesssim$ 0.5 $\mu$m in such a CDS of $\lesssim$ 40 
days, which is about the time between the first and third 
observation. However, to grow grains $>$ 0.5 $\mu$m within 
the same time through accretion higher densities are required. 
Alternatively, dust grains grow through coagulation rather than 
accretion.  

The above estimates of the cooling time of the postshock gas, 
the density of the CDS and mass of metals in the CDS are in 
good agreement with the requirements for rapid formation of 
dust. Comparison of the derived mass of metals to the inferred 
extinction and emission dust masses suggests a condensation 
efficiency of $\approx$ 1 $\%$ in the CDS. 

%
\section{Attaining large grain sizes and dust masses} 
Our estimated carbon dust mass of $\approx 2 \times 10^{-3}~\Msun$ 
at the late epoch (868 days) most likely represents a lower limit.
The inferred dust temperature from fits to the NIR emission 
allows for dust grain species other than carbon to be formed. 
Therefore, we explored different dust models (Methods) and found that 
either amorphous carbon or silicates 
can adequately fit the NIR emission. 
However, in either case, large grains provide good fits, while small carbon 
grains cannot be dominant.
For any model, the temperature range of the dust causing the NIR 
emission is between $\sim$ 1,000--1,300 K and the inferred amount 
of dust is for amorphous carbon $\sim$ 0.001--0.005 $\Msun$ 
and $\sim$ 0.006--0.023 $\Msun$ for silicate dust.

There is copious evidence of dust formation in the ejecta of 
Type IIP SNe (1, and references therein).
The theoretical models available in the literature are primarily
designed to model dust formation in the ejecta of 
SNe without a dense CSM, such as e.g., 
Type IIP or pair instability SNe 
\citesup[25,][]{2001MNRAS.325..726T, 2003ApJ...598..785N, 2007ApJ...666..955N, 
2007MNRAS.378..973B, 2010ApJ...713....1C},
or Type Ib and IIb SNe
\citesup{2008ApJ...684.1343N, 2010ApJ...713..356N}.
In these models, the density and temperature of the dust formation site 
(the ejecta) rapidly decline. Consequently, the time available for dust 
nucleation, which only happens in a certain range of high 
densities and moderate temperatures 
\citesup{1966AdPhy..15..111F, 1994LNP...428..163S},
is short. Subsequent grain growth at lower temperatures and densities 
becomes rapidly inefficient due to the fast declining densities. So far,
dust nucleation and growth can be accommodated up to 5 years after 
explosion (25), although grains grow to sizes 
no larger than (0.1--0.2 $\mu$m)
\citesup[25,][]{2003ApJ...598..785N}.
There are no dust formation models for luminous 
Type IIn SNe with a dense CSM, such as 
SN 2010jl (24).	
The SN ejecta, 
which sweeps up the dense CSM, could sustain higher 
densities over a longer period of time than predicted in the models 
without a CSM, leading to efficient grain growth and larger grains
as found in this paper. 

Alternatively, for ejecta dust formation, higher dust masses are inferred 
when invoking a clumpy structure of the dust forming region 
\citesup[17,][]{2006Sci...313..196S, 2007MNRAS.375..753E}.
Unfortunately the dust mass sensitively depends on the unknown optical depth 
of the clumps as well as the clump filling factors and distribution, making 
dust mass estimates uncertain. Nevertheless, a clumpy structure is possible
at the late epoch, even though it has been ruled out for the early dust 
formation in the CDS (Supplementary Information section~2).
There are no theoretical models addressing dust formation and growth 
in SNe featuring CSM -- ejecta interaction. More sophisticated radiative 
transfer modeling would be important to gain further insight into
dust mass and grain size estimates.

\newpage
%
\bibliographystylesup{naturemag}
\bibliographysup{sn2010jl}

\end{supp}
%
\end{document}